\documentclass[aps,prl,epsfig,graphics,floatfix,mathbbm,a4paper]{article}

\usepackage{amsmath,amsfonts,amssymb,graphics,graphicx,epsfig,color,times,bbm}

\newtheorem{question}{\indent Question}
\newtheorem{prop}{\indent Proposition}
\newtheorem{conj}{\indent Conjecture}
\newcommand{\lra}{\longrightarrow}

\newcommand{\C}{\mathbb{C}}
\newcommand{\G}{\mathcal{G}}
\newcommand{\E}{\mathcal{E}}
\newcommand{\1}{\mathbbm{1}}
\renewcommand{\>}{\rangle}
\newcommand{\<}{\langle}

 \DeclareMathOperator{\trace}{tr}

\DeclareMathOperator{\spanned}{span}
\DeclareMathOperator{\rank}{rank}
\DeclareMathOperator{\size}{size}

\newcommand{\proofend}{\hfill\fbox\\\medskip }

\begin{document}

%---------------------------------------------------------------------------

\newtheorem{theorem}{Theorem}
\newtheorem{proposition}{Proposition}

\newtheorem{lemma}{Proposition}

\newtheorem{definition}{Definition}
\newtheorem{corollary}{Corollary}

\newcommand{\proof}[1]{{\bf Proof.} #1 $\proofend$}
\vspace*{0.88truein}

\centerline{\bf
%%%%%%%%%%%%%%%%%%%%%
%Put in titiles here
%%%%%%%%%%%%%%%%%%%%%
MATRIX PRODUCT STATE REPRESENTATIONS} \vspace*{0.035truein}
\vspace*{0.37truein} \centerline{\footnotesize
%%%%%%%%%%%%%%%%%%%%%%%%%%%%%%%%%%%%
%put authors' name and address here
%%%%%%%%%%%%%%%%%%%%%%%%%%%%%%%%%%%%
D. PEREZ-GARCIA} \vspace*{0.015truein}
\centerline{\footnotesize\it Max Planck Institut f\"{u}r
Quantenoptik, Hans-Kopfermann-Str. 1, Garching, D-85748, Germany}
\baselineskip=10pt \centerline{\footnotesize\it Departamento de
An\'alisis Matem\'atico, Universidad Complutense de Madrid, 28040
Madrid, Spain} \vspace*{10pt}

\centerline{\footnotesize F. VERSTRAETE} \vspace*{0.015truein}
\centerline{\footnotesize\it Institute for Quantum Information,
Caltech, Pasadena, US} \baselineskip=10pt
\centerline{\footnotesize\it Fakult\"{a}t f\"{u}r Physik,
Universit\"{a}t Wien, Boltzmanngasse 5, A-1090 Wien, Austria.}
\vspace*{10pt}

\centerline{\footnotesize M.M. WOLF and J.I. CIRAC}
\vspace*{0.015truein} \centerline{\footnotesize\it Max Planck
Institut f\"{u}r Quantenoptik, Hans-Kopfermann-Str. 1, Garching,
D-85748, Germany} \vspace*{0.225truein}

%\publisher{(received date)}{(revised date)}

\vspace*{0.21truein}

%% \abstracts{first paragraph}{second paragraph}{third paragraph}
%% If there is only one paragraph, just keep the second and third empty
%% like the following one
\abstract{
%%%%%%%%%%%%%%%%%%%%
% put abstract here
%%%%%%%%%%%%%%%%%%%%
This work gives a detailed investigation of matrix product state
(MPS) representations for pure multipartite quantum states. We
determine the freedom in representations with and without
translation symmetry, derive respective canonical forms and
provide efficient methods for obtaining them. Results on
frustration free Hamiltonians and the generation of MPS are
extended, and the use of the MPS-representation for classical
simulations of quantum systems is discussed. }{}{}

\vspace*{10pt}

%%%%%%%%%%%%%%%%%%%%%%%%%%%%%%%%%%%%%%%%%%%%%%%%%%%%%%%%%%%%%%%%%%%%%%

%\begin{multicols}{2}

\tableofcontents

\section{Introduction and Overview}

The notorious complexity of quantum many-body systems stems to a
large extent from the exponential growth of the underlying Hilbert
space which allows for highly entangled quantum states. Whereas
this
 is a blessing for \emph{quantum information
theory}---it facilitates exponential speed-ups in quantum
simulation and quantum computing---it is often more a curse for
\emph{condensed matter theory} where the complexity of such
systems make them hardly tractable by classical means.
Fortunately, physical interactions are \emph{local} such that
states arising for instance as ground states from such
interactions are not uniformly distributed in Hilbert space.
Hence, it is desirable to have a representation of quantum
many-body states whose correlations are generated in a `local'
manner. Despite the fact that it is hard to make this picture
rigorous, there is indeed a representation which comes close to
this idea---the \emph{matrix product state} (MPS) representation.
In fact, this representation lies at the heart of the power of the
\emph{density matrix renormalization group} (DMRG) method and it
is the basis for a large number of recent developments in quantum
information as well as in condensed matter theory.

This work gives a detailed investigation of the MPS representation
with a particular focus on the freedom in the representation and
on canonical forms. The core of our work is a generalization of
the results on finitely correlated states in \cite{FaNaWe92} to
finite systems with and without translational invariance. We will
mainly discuss exact MPS representations throughout and just
briefly review results on approximations in
Sec.\ref{sec:classical-sim}. In order to provide a more complete
picture of the representation and its use we will also briefly
review and extend various recent results based on MPS, their
parent Hamiltonians and their generation. The following gives an
overview of the article and sketches the obtained results:
\begin{itemize}
    \item Sec.\ref{Sec:prelim} will introduce the basic notions,
    provide some examples
    and give an overview over the relations between MPS and the valence bond picture on the one
    hand and frustration free Hamiltonians and finitely correlated
    states on the other.
    \item In Sec.\ref{Sec:canonical} we will determine the freedom
    in the MPS representation, derive canonical forms and provide
    efficient ways for obtaining them. Cases with and without
    translational invariance are distinguished. In the former
    cases we show that there is always a translational invariant representation and derive a canonical decompositions of states into superpositions
    of `ergodic' and periodic states (as in \cite{FaNaWe92}).
    \item Sec.\ref{Sec:Hamiltonians} investigates a standard
    scheme which constructs for any MPS a local Hamiltonian, which
    has the MPS as exact ground state. We prove uniqueness of the
    ground state (for the generic case) without referring to the
    thermodynamic limit, discuss degeneracies (spontaneous symmetry breaking) based on the canonical decomposition and review results on uniform bounds to
    the energy gap.
    \item In Sec.\ref{sec:generation} we will review the connections between MPS and
    sequential generation of multipartite entangled states. In particular we will show
    that MPS of sufficiently small bond dimension are feasible to generate in a lab.
    \item In Sec.\ref{sec:classical-sim} we will review the results that show how MPS efficiently
    approximate many important states in nature; in particular, ground states of 1D
    local Hamiltonians. We will also show how the MPS formalism is crucial to
    understand the need of a large amount of entanglement in a quantum computer in
    order to have a exponential speed-up with respect to a classical one.
\end{itemize}

\section{Definitions and Preliminaries}\label{Sec:prelim}
\subsection{MPS and the valence bond picture}
We will throughout consider pure quantum states $|\psi\rangle\in
\mathbb{C}^{\otimes d^N}$ characterizing a system of $N$ sites
each of which corresponds to a $d$-dimensional Hilbert space. A
very useful and intuitive way of thinking about MPS is the
following valence bond construction: consider the $N$ parties
('spins') aligned on a ring and assign two virtual spins of
dimension $D$ to each of them. Assume that every pair of
neighboring virtual spins which correspond to different sites are
initially in an (unnormalized) maximally entangled state
$|I\rangle=\sum_{\alpha=1}^{D} |\alpha,\alpha\rangle$ often
referred to as entangled \emph{bond}. Then apply a map
\begin{equation}\label{curlyA}{\cal
A}=\sum_{i=1}^d\sum_{\alpha,\beta=1}^{D} A_{i,\alpha,\beta}
|i\rangle\langle\alpha,\beta|\end{equation} to each of the $N$
sites. Here and in the following Greek indices correspond to the
virtual systems. By writing $A_{i}$ for the $D\times D$ matrix
with elements $A_{i,\alpha,\beta}$ we get that the coefficients of
the final state when expressed in terms of a product basis are
given by a matrix product $\trace\left[A_{i_1}A_{i_2}\cdots
A_{i_N}\right]$. In general the dimension of the entangled state
$|I\rangle$ and the map ${\cal A}$ can both be site-dependent and
we write $A_i^{[k]}$ for the $D_k\times D_{k+1}$ matrix
corresponding to site $k\in\{1,\ldots,N\}$. States obtained in
this way have then the form
\begin{equation}\label{MPS0}
|\psi\rangle=\sum_{i_1,\ldots,i_N=1}^d
\trace\left[A_{i_1}^{[1]}A_{i_2}^{[2]}\cdots
A_{i_N}^{[N]}\right]|i_1,i_2,\ldots,i_N\rangle\;,
\end{equation}
and are called \emph{matrix product states} \cite{MPSorigin}. As
shown in \cite{Vi03} \emph{every} state can be represented in this
way if only the bond dimensions $D_k$ are sufficiently large.
Hence, Eq.(\ref{MPS0}) is a representation of states rather than
the characterization of a specific class. However, typically
states are referred to as  MPS if they have a MPS-representation
with small $D=\max_k D_k$ which (in the case of a sequence of
states) does in particular not grow with $N$. Note that $\psi$ in
Eq.(\ref{MPS0}) is in general not normalized and that its MPS
representation is not unique. Normalization as well as other
expectation values of product operators can be obtained from
\begin{eqnarray}\nonumber \langle\psi|\bigotimes_{k=1}^N
S_k|\psi\rangle &=&\trace\left[\prod_{k=1}^N
E_{S_k}^{[k]}\right]\;,\quad\text{with}\\
E_S^{[k]}&\equiv&\sum_{i,j=1}^d \langle i|S|j\rangle
\overline{A}_i^{[k]}\otimes A_j^{[k]}\label{E}\;.\end{eqnarray}

\begin{figure} [htbp]
%\vspace*{13pt}
\centerline{\epsfig{file=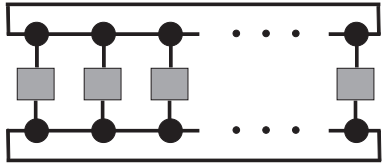, width=5cm}} %100 percent
\vspace*{13pt} \caption{\label{tensor-net}{\small Computing an
expectation value of an MPS is equivalent to contract the tensor
  of the figure, where bonds represent indices that are contracted. The matrices associated to each spin are represented
  by the circles (the vertical bond of each matrix is its physical index) and observables are represented by squares.
  It is trivial to see that this contraction can be done efficiently.}}
\end{figure}

\subsection{Finitely correlated states}
The present work is inspired by the papers on \emph{finitely
correlated states} (FCS) which in turn generalize the findings of
Affleck, Kennedy, Lieb and Tasaki (AKLT) \cite{AKLT}. In fact,
many of the results we derive are extensions of the FCS formalism
to finite and/or non-translational invariant systems. For this
reason we will briefly review the work on FCS. A FCS is a
translational invariant state on an infinite spin chain which is
constructed from a completely positive and trace preserving map
$\mathbb{E}:{\cal B}({\cal H}_A)\rightarrow {\cal B}({\cal
H}_A\otimes {\cal H}_B)$ and a corresponding fixed point density
operator $\Lambda=\trace_B[\mathbb{E}(\Lambda)]$. Here ${\cal
H}_B=\mathbb{C}^d$ is the Hilbert space corresponding to one site
in the chain and ${\cal H}_A=\mathbb{C}^D$ is an ancillary system.
An $n$-partite reduced density matrix $\rho_n$ of the FCS is then
obtained by repeated application of $\mathbb{E}$ to the ancillary
system (initially in $\Lambda$) followed by tracing out the
ancilla, i.e., \begin{equation}
\rho_n=\trace_A\big[\mathbb{E}^n(\Lambda)\big]\;.
\end{equation}
An important instance are \emph{purely generated} FCS where
$\mathbb{E}(x)=V^\dag x V$ is given by a partial isometry $V$. The
latter can be easily related to the $A$'s in the matrix product
representation via $V=\sum_{i=1}^d\sum_{\alpha,\beta=1}^D
A_{i,\alpha,\beta} |\alpha\rangle\langle\beta i|$. Expressed in
terms of the matrices $A_i$ the isometry condition and the fixed
point relation  read \begin{equation}\label{FCSA} \sum_{i=1}^d
A_iA_i^\dag=\mathbbm{1}\;,\qquad \sum_{i=1}^d A_i^\dag\Lambda A_i
=\Lambda\;,
\end{equation} which already anticipates the type of canonical forms for
MPS discussed below. As shown in \cite{FaNaWe92-2} purely
generated FCS are weakly dense within the set of all translational
invariant states on the infinite spin chain. Moreover, a FCS is
\emph{ergodic}, i.e., an extreme point within all translational
invariant states, iff the map $\E(x)=\sum_iA_ixA_i^\dag$ has a
non-degenerate eigenvalue 1 (i.e., $\mathbbm{1}$ and $\Lambda$ are
the only fixed points in Eq.(\ref{FCSA})). Every FCS has a unique
decomposition into such ergodic FCS which in turn can be
decomposed into $p$ $p$-periodic states each of which corresponds
to a root of unity $\exp(\frac{2\pi i}p m)$, $m=0,\ldots,p-1$ in
the spectrum of $\E$. A FCS is pure iff it is purely generated and
1 is the only eigenvalue of $\E$ of modulus 1. In this case the
state is \emph{exponentially clustering}, i.e., the connected
two-point correlation functions decay exponentially
\begin{equation}\label{eq:corr}
\langle S_i\otimes\mathbbm{1}^{\otimes l-1}\otimes
S_{i+l}\rangle-\langle S_i\rangle\langle S_{i+l}\rangle={\cal
O}\big(|\nu_2|^{l-1}\big)\;,
\end{equation} where $\nu_2$ ($|\nu_2|<1$) is the second largest eigenvalue of
$\E$.

\newpage\subsection{Frustration free Hamiltonians} Consider a
translational invariant Hamiltonian on a ring of $N$
$d$-dimensional quantum systems
\begin{equation}H=\sum_{i=1}^N \tau^i\big(h\big)\;,
\end{equation}
where $\tau$ is the translation operator with periodic boundary
conditions, i.e., $\tau\big(\bigotimes_{i=1}^N
x_i\big)=\bigotimes_{i=1}^N x_{i+1}$ where sites $N+1$ and $1$ are
identified. The interaction is called $L$-\emph{local} if $h$ acts
non-trivially only on $L$ neighboring sites, and it is said to be
\emph{frustration free} with respect to its ground state $\phi_0$
if the latter minimizes the energy locally in the sense that
$\langle \phi_0|H|\phi_0\rangle=\inf_\phi \langle
\phi|H|\phi\rangle=N \inf_\phi \langle \phi|h|\phi\rangle$. As
proven in \cite{Hast1} all gapped Hamiltonians can be approximated
by frustration free ones if one allows for enlarging the
interaction range $L$ up to ${\cal O}(\log N)$.

For every MPS and FCS $\psi$ one can easily find frustration free
Hamiltonians such that $\psi$ is their exact ground state.
Moreover, these \emph{parent} Hamiltonains are $L$-local with
$L\sim 2 \log D/\log d$ and they allow for a detailed analysis of
the ground state degeneracy (Sec.\ref{sec:Huniqueness}) and the
energy gap above the ground state (Sec.\ref{Sec:Hgap}). Typically,
these Hamiltonians are, however, not exactly solvable, i.e.,
information about the excitations might be hard to obtain.

\subsection{Examples}
\begin{enumerate}
    \item {\it AKLT}: The father of all matrix product states is
    the ground state of the AKLT-Hamiltonian
    \begin{equation}\label{HAKLT}
    H=\sum_i
    \vec{S}_i\vec{S}_{i+1}+\frac13\Big(\vec{S}_i\vec{S}_{i+1}\Big)^2\;,
    \end{equation}
    where $\vec{S}$ is the vector of spin-1 operators (i.e., d=3).
    Its MPS representation is given by
    $\{A_i\}=\big\{\sigma^z,\sqrt{2}\sigma^+,-\sqrt{2}\sigma^-\big\}$
    where the $\sigma$'s are the Pauli matrices.
    \item {\it Majumdar-Gosh}: The  Hamiltonian
    \begin{equation}\label{HMG}
    H=\sum_i 2\vec{\sigma}_i\vec\sigma_{i+1} +
    \vec{\sigma}_i\vec\sigma_{i+2}\;
    \end{equation}
    is such that every ground state is a superposition of two 2-periodic states given by products of singlets on neighboring sites.
    The equal weight superposition of these states is
    translational invariant and has an MPS representation
    \begin{equation}
    A_1=\left(%
\begin{array}{ccc}
  0 & 1 & 0 \\
  0 & 0 & -1 \\
  0 & 0 & 0 \\
\end{array}%
\right)\;,\quad A_2=\left(\begin{array}{ccc}
  0 & 0& 0 \\
  1 & 0 & 0 \\
  0 & 1 & 0 \\
\end{array}%
\right)\;.
    \end{equation}
    \item {\it GHZ states} of the form $|\psi\rangle=|++\ldots +\rangle+|--\ldots
    -\rangle$ have an MPS representation $A_\pm=\mathbbm{1}\pm\sigma^z$. Anti-ferromagnetic GHZ states would correspond to  $A_\pm=\sigma^\pm$.
    \item {\it Cluster states} are unique ground states of the
    three-body interactions $\sum_i
    \sigma^z_i\sigma^x_{i+1}\sigma^z_{i+2}$ and represented by the
    matrices $$ A_1= \left(%
\begin{array}{cc}
  0 & 0 \\
  1 & 1 \\
\end{array}%
\right)\;,\quad A_2=\left(%
\begin{array}{cc}
  1 & -1 \\
  0 & 0 \\
\end{array}%
\right)\;.$$
    \item {\it W-states} can for instance appear as ground states of the ferromagnetic
    XX model with strong transversal magnetic field. A W-state is
    an equal superposition of all translates of
    $|100\ldots00\rangle$. For a simple MPS representation choose $\{A_1^{[k]},A_2^{[k]}\}$ equal to $\{\sigma^+,\mathbbm{1}\}$
    for all $k< N$ and $\{\sigma^+\sigma^x,\sigma^x\}$ for $k=N$. Although the state itself is translational invariant
    there is no MPS representation with $D=2$ having this
    symmetry.
\end{enumerate}

\section{The canonical form}\label{Sec:canonical}

The general aim of this section will be to answer the following
questions about the MPS representation of a given pure state:

\begin{question}\label{quest.can.1}
Which is the freedom in the representation? \end{question}
\begin{question}\label{quest.can.2} Is there any  {\it canonical}
representation? \end{question}
\begin{question}\label{quest.can.3}
If so, how to get it?
\end{question}

We will distinguish two cases. The general case, or the case of
open boundary conditions (OBC) and the case in which one has the
additional properties of translational invariance (TI) and
periodic boundary conditions (PBC).

\subsection{Open boundary conditions}\label{sec:Open}

A MPS is said to be written with open boundary conditions (OBC) if
the first and last matrices are vectors, that is, if it has the
form
\begin{equation}\label{eq.vidal}
|\psi\rangle=\sum_{i_1,\ldots,i_N}
A^{[1]}_{i_1}A^{[2]}_{i_2}\cdots
A^{[N-1]}_{i_{N-1}}A^{[N]}_{i_N}|i_1\cdots i_N\rangle,
\end{equation}
where $A^{[m]}_i$ are $D_{m}\times D_{m+1}$ matrices with
$D_1=D_{N+1}=1$. Moreover, if $D=\max_m D_m$ we say that the MPS
has {\it (bond) dimension} $D$. The following is shown in
\cite{Vi03}:
\begin{theorem}[Completeness and canonical form]\label{thm:OBC-Vidal} Any
state $\psi\in\mathbb{C}^{d\otimes N}$ has an OBC-MPS
representation of the form Eq.(\ref{eq.vidal}) with bond dimension
$D\leq d^{\lfloor N/2\rfloor}$ and
\begin{enumerate}
\item $\sum_i A^{[m]}_iA^{[m]\dagger}_i=\mathbbm {1}_{D_m}$ for
all $1\le m\le N$. \item $\sum_i A^{[m]\dagger}_i\Lambda^{[m-1]}
A^{[m]}_i=\Lambda^{[m]}, $ for all $1\le m\le N$,  \item
$\Lambda^{[0]}=\Lambda^{[N]}=1$ and each $\Lambda^{[m]}$ is a
$D_{m+1}\times D_{m+1}$ diagonal matrix which is positive, full
rank and with $\trace{\Lambda^{[m]}}=1$.
\end{enumerate}\label{Thm:completeness}
\end{theorem}
Thm.\ref{Thm:completeness} is proven by successive singular value
decompositions (SVD), i.e., Schmidt decompositions in $\psi$, and
the \emph{gauge conditions} {\it 1.-3.} can be imposed by
exploiting the simple observation that
$A_i^{[m]}A_j^{[m+1]}=(A_i^{[m]}X) (X^{-1}A_j^{[m+1]})$. If {\it
1.-3.} are satisfied for a MPS representation, then we say that
the MPS with OBC is in {\it the canonical form}. From the way it
has been obtained one immediately sees that:
\begin{itemize}
\item  it is unique (up to permutations and degeneracies in the
Schmidt Decomposition), \item $\Lambda^{[m]}$ is the diagonal
matrix of the non-zero eigenvalues of the reduced density operator
$\rho_m=\trace_{m+1,\ldots,N}|\psi\rangle\langle\psi|$, \item  any
state for which $\max_m \rank(\rho_m)\le D$ can be written as a
MPS of bond dimension D.
\end{itemize}

This answers questions \ref{quest.can.2} and \ref{quest.can.3}.
Question \ref{quest.can.1} will be answered with the next theorem
which shows that the entire freedom in any OBC-MPS representation
is given by `local' matrix multiplications.

\begin{theorem}[Freedom in the choice of the matrices]\label{free-OBC}
Let us take a OBC-MPS representation \begin{equation*}
|\psi\rangle=\sum_{i_1,\ldots,i_N}
B^{[1]}_{i_1}B^{[2]}_{i_2}\cdots
B^{[N-1]}_{i_{N-1}}B^{[N]}_{i_N}|i_1\cdots i_N\rangle\;.
\end{equation*}
Then, there exist (in general non-square) matrices $Y_j$, $Z_j$
with  $Y_jZ_j=\mathbbm{1}$ such that, if we define
\begin{align}
\nonumber A^{[1]}_i&=B^{[1]}_iZ_1,\quad  A^{[N]}_i =Y_{N-1}B^{[N]}_i \\
A^{[m]}_i &=Y_{m-1}B^{[m]}_iZ_m,\ \text{for}\ 1<m<N
\label{eq.lem.II.9}
%\nonumber A^{[N-1]}_i &=Y_{N-2}B^{[N-1]}_iZ_{N-1} \\
\end{align}
the canonical form is given by
\begin{equation}\label{Eq:canonical2}
|\psi\rangle=\sum_{i_1,\ldots,i_N}
A^{[1]}_{i_1}A^{[2]}_{i_2}\cdots
A^{[N-1]}_{i_{N-1}}A^{[N]}_{i_N}|i_1\cdots i_N\rangle.
\end{equation}
\end{theorem}

\proof{We will prove the theorem in three steps.

\textbf{\it{STEP 1.}}  First we will find the matrices $A^{[j]}_i$
verifying relation (\ref{eq.lem.II.9}) but just with the property
$\sum_i A^{[j]}_iA^{[j]\dagger}_i=\mathbbm{1}.$

To this end we start from the right by doing SVD:
$B^{[N]}_{\alpha,i}=\sum_{\beta}U^{[N-1]}_{\alpha,\beta}\Delta^{[N-1]}_{\beta}
A^{[N]}_{\beta,i}$, with $U^{[N-1]},A^{[N]}$ unitaries and
$\Delta^{[N-1]}$ diagonal. That is $B^{[N]}_i=Z_{N-1}A_i^{[N]}$,
with $Z_{N-1}=U^{[N-1]}\Delta^{[N-1]}$. Clearly
$\sum_iA_i^{[N]}A_i^{[N]\dagger}=\mathbbm{1}$ and $Z_{N-1}$ has a
left inverse. Now we call $\tilde{B}^{[N-1]}_i=B^{[N-1]}_iZ_{N-1}$
and make another SVD:
$\tilde{B}^{[N-1]}_{\alpha,i,\beta}=\sum_\gamma
U^{[N-2]}_{\alpha,\gamma}\Delta^{[N-2]}_\gamma
A^{[N-1]}_{\gamma,i,\beta}$. That is
$$B^{[N-1]}_iZ_{N-1}=\tilde{B}^{[N-1]}_i=Z_{N-2}A_i^{[N-1]},$$
where $\sum_iA_i^{[N-1]}A_i^{[N-1]\dagger}=\mathbbm{1}$ and
$Z_{N-2}=U^{[N-2]}\Delta^{[N-2]}$ has left inverse.

We can go on getting relations (\ref{eq.lem.II.9}) to the last
step, where one simply defines $A^{[1]}_i=B^{[1]}_iZ_1$. From the
construction one gets Eq.(\ref{Eq:canonical2})
%\begin{equation*}
%|\psi\rangle=\sum_{i_1,\ldots,i_N}
%^{[1]}_{i_1}A^{[2]}_{i_2}\cdots
%A^{[N-1]}_{i_{N-1}}A^{[N]}_{i_N}|i_1\cdots i_N\rangle.
%\end{equation*}
and that $\sum_i A^{[j]}_iA^{[j]\dagger}_i=\mathbbm{1}$ for every
$1<j\le N$. The case $j=1$ comes simply from the normalization of
the state: $$1=\<\psi|\psi\>=\sum_{i_1,\ldots,i_N}
A^{[1]}_{i_1}\cdots A^{[N]}_{i_N}A^{[N] \dagger}_{i_N}\cdots
A^{[1] \dagger}_{i_1}=\sum_{i_1}A^{[1]}_{i_1}A^{[1]
\dagger}_{i_1},$$ where in the last equality we have used that
$\sum_{i_j}A^{[j]}_{i_j}A^{[j] \dagger}_{i_1}=\1$ for $1<j\le N$.

\textbf{\it{STEP 2.}} Now we can assume that the $B$'s verify
$\sum_i B^{[j]}_iB^{[j]\dagger}_i=\mathbbm{1}$. Diagonalizing
$\sum_i B^{[1]\dagger}_iB^{[1]}_i$ we get a unitary $V^{[1]}$ and
a positive diagonal matrix $\Lambda^{[1]}$ such that $\sum_i
B^{[1]\dagger}_iB^{[1]}_i=V^{[1]}\Lambda^{[1]}V^{[1]\dagger}$.
Calling $A^{[1]}_i=B^{[1]}_iV^{[1]}$ we have both
$\sum_iA_i^{[1]}A_i^{[1]\dagger}=\mathbbm{1}$ and
$\sum_iA_i^{[1]\dagger}A_i^{[1]}=\Lambda^{[1]}$.

Now we diagonalize $\sum_i
B^{[2]\dagger}_iV^{[1]}\Lambda^{[1]}V^{[1]\dagger}B^{[2]}_i=V^{[2]}\Lambda^{[2]}V^{[2]\dagger}$
and define $A^{[2]}_i=V^{[1]\dagger}B^{[2]}_iV^{[2]}$ to have both
$\sum_iA_i^{[2]}A_i^{[2]\dagger}=\mathbbm{1}$ and
$\sum_iA_i^{[2]\dagger}\Lambda^{[1]}A_i^{[2]}=\Lambda^{[2]}$. We
keep on with this procedure to the very last step where we simply
define $A_i^{[N]}=V^{[N-1]\dagger}B^{[N]}_i$.
$\sum_iA_i^{[N]}A_i^{[N]\dagger}=\mathbbm{1}$  is trivially
verified and
$\sum_iA_i^{[N]\dagger}\Lambda^{[N-1]}A_i^{[N]}=\Lambda^{[N]}=1$
comes, as above, from the normalization of the state. Moreover, by
construction we have the relation (\ref{eq.lem.II.9}) and
Eq.(\ref{Eq:canonical2}).
%\begin{equation*} |\psi\rangle=\sum_{i_1,\ldots,i_N}
%A^{[1]}_{i_1}A^{[2]}_{i_2}\cdots
%A^{[N-1]}_{i_{N-1}}A^{[N]}_{i_N}|i_1\cdots i_N\rangle.
%\end{equation*}

\textbf{\it{STEP 3.}} At this point we have matrices ${Y}_j,
{Z}_j$ with ${Y}_jZ_j=\1$ such that, if we define ${A}_i^{[j]}$ by
(\ref{eq.lem.II.9}), we get ${D}_j\times {D}_{j+1}$ matrices
verifying the conditions 1, 2 and 3 of Theorem \ref{thm:OBC-Vidal}
with the possible exception that the matrices $\Lambda^{[j]}$ are
not full rank. Now we will show that we can redefine $Y_j, Z_j$
(and hence $D_j, A_i^{[j]}$) to guarantee also this {\it full
rank} condition.

We do it by induction. Let us assume that $\Lambda^{[j-1]}$ is
full rank and the $D_{j+1}\times D_{j+1}$ positive diagonal matrix
$\Lambda^{[j]}$ is not. Then, calling
$$\widetilde{D}_{j+1}=\rank(\Lambda^{[j]})\;, \quad
P_j=\left(\1_{\widetilde{D}_{j+1}}\big|0_{D_{j+1}-\widetilde{D}_{j+1}}\right),$$
we are finished if we update $Z_j$ as $Z_jP_j^{\dagger}$, $Y_j$ as
$P_jY_j$ (and hence ${D}_{j+1}$ as $\tilde{D}_{j+1}$, $A_i^{[j]}$
as $A_i^{[j]}P_j^{\dagger}$, $A_i^{[j+1]}$ as $P_jA_i^{[j+1]}$ and
$\Lambda^{[j]}$ as $P_j\Lambda^{[j]}P_j^{\dagger}$). The only
non-trivial part is to prove that
$A^{[j]}_{i_j}A^{[j+1]}_{i_{j+1}}=A^{[j]}_{i_j}P_j^{\dagger}P_jA^{[j+1]}_{i_{j+1}}$.
For that, calling $C=\1_{D_{j+1}}-P_j^{\dagger}P_j$, it is
enough to show that $A^{[j]}_iC=0$. Since $$\1_{D_{j+1}}-P_j^{\dagger}P_j=\left(%
\begin{array}{cc}
  0_{\widetilde{D}_{j+1}} & 0 \\
  0 &  \1_{D_{j+1}-\widetilde{D}_{j+1}} \\
\end{array}%
\right), $$  we have
$$0=C\Lambda^{[j]}C=\sum_iC
A^{[j]\dagger}_i\Lambda^{[j-1]}A^{[j]}_i C.$$ Since
$\Lambda^{[j-1]}$ is positive and full rank we get $A^{[j]}_iC=0$.
}

%\section{Our contribution: Periodic Boundary Conditions and Translational
%Invariance}

\subsection{Periodic boundary conditions and translational invariance}

Clearly, if the $A's$ in the MPS in Eq.(\ref{MPS0}) are the same,
i.e., site-independent ($A_i^{[m]}=A_i$), then the state is
translationally invariant (TI) with periodic boundary conditions
(PBC). We will in the following first show that the converse is
also true, i.e., that every TI state has a TI MPS representation.
Then we will derive canonical forms having this symmetry, discuss
their properties and show how to obtain them. An important point
along these lines will be a canonical decomposition of TI states
into superpositions of TI MPS states which may in turn be written
as superpositions of periodic states. This decomposition closely
follows the ideas of \cite{FaNaWe92} and will later, when
constructing parent Hamiltonians, give rise to discrete
symmetry-breaking.

\subsubsection{Site independent matrices}

Before starting with the questions \ref{quest.can.1},
\ref{quest.can.2} and \ref{quest.can.3}, we will see that we can
use TI and PBC to assume the matrices in the MPS representation to
be site independent. That is, if the state is TI, then there is
also a TI representation as MPS.\footnote{In a similar way other
symmetries can be restored in the representation. For example if
the state is reflection symmetric then we can find a
representation with $A_i=A_i^T$ and if it is real in some basis
then we can choose one with real $A_i$. Both representations are
easily obtained by doubling the bond dimension $D$. For the
encoding of other symmetries in the $A$'s we refer to
\cite{FaNaWe92,NachtergaeleSym,Ortiz}.}

\begin{theorem}[Site-independent matrices]
{\it Every} TI pure state with PBC on a finite chain has a MPS
representation with site-independent matrices $A_i^{[m]}=A_i$,
i.e.,\begin{equation}\label{eq.1}
|\psi\rangle=\sum_{i_1,\ldots,i_N} \trace(A_{i_1}\cdots
A_{i_N})|i_1\cdots i_N\rangle\;.
\end{equation}
If we start from an OBC MPS representation, to get
site-independent matrices one has (in general) to increase the
bond dimension from $D$ to $ND$ (note the $N$-dependence).
\end{theorem}

\proof{ We start with an OBC representation of the state with
site-dependent $A_i^{[m]}$
%$$|\psi\rangle=\sum_{i_1,\ldots,i_N=0}^{d-1}A^{[1]}_{i_1}\cdots A^{[N]}_{i_N}|i_1,\ldots,i_N\rangle.$$
and consider the matrices (for $0\le i\le d-1$)
    $$B_i=N^{-\frac1N}\begin{pmatrix}
      0 & A^{[1]}_i &  &  &  \\
       & 0 & A^{[2]}_i &  &  \\
       &  & \cdots &  &  \\
     &  &  & 0 & A^{[N-1]}_i \\
      A^{[N]}_i &  & &  & 0
    \end{pmatrix}.$$
This leads to
           $$\sum_{i_1,\ldots,i_N=0}^{d-1}\trace(B_{i_1}\cdots
        B_{i_N})|i_1,\ldots,i_N\rangle=$$
        $$=\frac1N\sum_{j=0}^{N-1}\sum_{i_1,\ldots,i_N=0}^{d-1}\trace(A^{[1]}_{i_{1+j}}\cdots A^{[N]}_{i_{N+j}})
        |i_1,\ldots,i_N\rangle,$$
where $i_{j}=i_{j-N}$ if $j>N$. Due to TI of $\psi$ this yields
exactly Eq.(\ref{eq.1}). }

To explicitly show the $N$-dependence of the above construction we
consider the particular case of the $W$-state
$|10\ldots0\>+|01\ldots 0\> +\cdots+|0\ldots 01\>$. In this case
the minimal bond dimension is $2$ as a MPS with OBC. However, if
we want site-independent matrices, it is not difficult to show
that one needs bigger matrices. In fact, we conjecture that the
size of the matrices has to grow with $N$ (Appendix
\ref{open.problems}).

From now on we suppose that we are dealing with a MPS of the form
in Eq.(\ref{eq.1})
%\begin{equation*}
%|\psi\rangle=\sum_{i_1,\ldots,i_N} \trace(A_{i_1}\cdots
%A_{i_N})|i_1\cdots i_N\rangle,
%\end{equation*}
with the matrices $A_i$ of size $D\times D$. In cases where we
want to emphasize the site-independence of the matrices, we say
the state is TI represented or simply a  TI MPS.
%, which will be called
%a MPS of (bond) dimension $D$, or simply a $D$-MPS.

\subsubsection{MPS and CP maps}

There is a close relation (and we will repeatedly use it) between
a TI MPS
%\begin{equation*}
%|\psi\rangle=\sum_{i_1,\ldots,i_N} \trace(A_{i_1}\cdots
%A_{i_N})|i_1\cdots i_N\rangle,
%\end{equation*}
and the completely positive map $\E$ acting on the space of
$D\times D$ matrices given by
\begin{equation}\label{eq.E}
\E(X)=\sum_iA_iXA_i^{\dagger}.
\end{equation}
One can always assume without loss of generality that the cp map
$\E$ has spectral radius equal to $1$ which implies by
\cite[Theorem 2.5]{EvansH-Krohn78} that \textit{$\E$ has a
positive fixed point}. As in the FCS case stated in
Eq.(\ref{eq:corr}) the second largest eigenvalue of $\E$
determines the correlation length of the state and as we will see
below the eigenvalues of magnitude one are closely related to the
terms in the canonical decomposition of the state. Note that $\E$
and $E_\mathbbm{1}=\sum_i A_i\otimes \bar{A}_i$ have the same
spectrum as they are related via \begin{equation}\label{E2E}
\langle \beta_1|\E(|\alpha_1\rangle\langle
\alpha_2|)|\beta_2\rangle
=\langle\beta_1,\beta_2|E_\mathbbm{1}|\alpha_1,\alpha_2\rangle\;.
\end{equation}

Since the Kraus operators of the cp map $\E$ are uniquely
determined up to unitaries, it implies that $\E$ uniquely
determines the MPS up to local unitaries in the physical system.
This is used in \cite{renorm} to find the fixed points of a
renormalization group procedure on quantum states. There it is
made explicit in the case of qubits, where a complete
classification of the cp-maps is known. To be able to characterize
the fixed points in the general case one has to find the reverse
relation between MPS and cp maps. That is, given a MPS, which are
the possible $\E$ that can arise from different matrices in the
MPS representation? It is clear that a complete solution to
question \ref{quest.can.1} will give us the answer. However,
though we will below provide the answer in the generic case, this
is far from being completely general. As a simple example of how
different the cp-maps $\E$ can be for the {\it same} MPS, let us
take an arbitrary cp-map $\E(X)=\sum_iA_iXA_i^{\dagger}$ and
consider the associate MPS for the case of $2$ particles:
$|\psi\rangle=\sum_{i_1,i_2} \trace(A_{i_1}A_{i_2})|i_1
i_2\rangle$. Now translational invariance means permutational
invariance and hence it is not difficult to show that there exist
diagonal matrices $D_i$ such that $|\psi\rangle=\sum_{i_1,i_2}
\trace(D_{i_1}D_{i_2})|i_1 i_2\rangle$. This defines a new cp-map
$\tilde{\E}(X)=\sum_iD_iXD_i^{\dagger}$ with diagonal Kraus
operators, for which e.g many of the additivity conjectures are
true \cite{diagonal-krauss}.

 \

\subsubsection{The canonical representation}

In this section we will show that one can always decompose the
matrices of a TI-MPS to a {\it canonical} form. Subsequently we
will discuss a generic condition based on which the next section
will answer question \ref{quest.can.2} concerning the uniqueness
of the canonical form.

\begin{theorem}[TI canonical form]\label{Th:TIcanonical} Given a TI state
on a finite ring, we can {\it always} decompose the matrices $A_i$
of any of its TI MPS representations as
$$A_i=\left(%
\begin{array}{ccc}
  \lambda_1 A_i^1 & 0 & 0 \\
  0 & \lambda_2 A_i^2 & 0 \\
  0 & 0 & \cdots \\
\end{array}%
\right),$$ where $1\ge \lambda_j>0$ for every $j$ and the matrices
$A_i^j$ in each block verify the conditions:
\begin{enumerate}
\item $\sum_i A_i^jA_i^{j\dagger}=\mathbbm {1}$. \item $\sum_i
A_i^{j\dagger}\Lambda^j A_i^j=\Lambda^j,$  for some diagonal
positive and full-rank matrices $\Lambda^j$. \item $\mathbbm {1}$
is the only fixed point of the operator $\E_j(X)=\sum_i
A_i^jXA_i^{j\dagger}$.
\end{enumerate}

If we start with a TI MPS representation with bond dimension $D$,
the bond dimension of the above {\em canonical form} is $\le D$.
\end{theorem}

\proof{ We assume w.l.o.g. that the spectral radius of $\E$ is $1$
(this is where the $\lambda_j$ appear) and we denote by $X$ a
positive fixed point of $\E$. If $X$ is invertible, then calling
$B_i=X^{-\frac{1}{2}}A_i X^{\frac{1}{2}}$ we have $\sum_i
B_iB_i^{\dagger}=\mathbbm{1}$ and hence condition 1.

If $X$ is not invertible and we write $X=\sum_\alpha
\lambda_{\alpha} |\alpha\>\<\alpha|$, and we call $P_{R}$ the
projection onto the subspace $R$ spanned by the $|\alpha\>$'s,
then we have that $A_iP_R=P_RA_iP_R$ for every $i$. To see this,
it is enough to show that $A_i|\alpha\>\in R$ for every
$i,|\alpha\>$. If this does not happen for some $j,|\beta\>$, then
$\sum_\alpha \lambda_{\alpha} |\alpha\>\<\alpha|- \lambda_\beta
A_j|\beta\>\<\beta|A_j^{\dagger}\not \ge 0$. But, since
$\sum_\alpha \lambda_{\alpha} |\alpha\>\<\alpha|=\sum_i
\sum_\alpha \lambda_{\alpha} A_i|\alpha\>\<\alpha|A_i^{\dagger}$,
we have obtained that $$\sum_{(i,\alpha)\not =(j,\beta)}
\lambda_{\alpha} A_i|\alpha\>\<\alpha|A_i^{\dagger}\not \ge 0,$$
which is the desired contradiction.

If we call $R^{\perp}$  the orthogonal subspace of $R$, we can
decompose our state as
\begin{eqnarray}
|\psi\>&=&\sum_{i_1,\ldots,i_N} \trace_R(A_{i_1}\cdots
A_{i_N})|i_1\cdots i_N\rangle+\nonumber\\ &&+\sum_{i_1,\ldots,i_N}
\trace_{R^{\perp}}(A_{i_1}\cdots A_{i_N})|i_1\cdots
i_N\rangle.\nonumber
\end{eqnarray}
On the one hand $\trace_R(A_{i_1}\cdots A_{i_N})$ is given by
$$ \trace(P_RA_{i_1}\cdots
A_{i_N}P_R)=  \trace(P_RA_{i_1}P_R\cdots P_RA_{i_N}P_R)$$ which
corresponds to a MPS with matrices $B_i=P_RA_iP_R$ of size
$\dim(R)\times \dim(R)$. On the other hand
\begin{eqnarray}
\nonumber \trace_{R^{\perp}}(A_{i_1}\cdots
A_{i_N})&=&\trace(P_{R^{\perp}}A_{i_1}\cdots A_{i_N}P_{R^{\perp}})\\
\nonumber &=& \trace(P_{R^{\perp}}A_{i_1}P_{R^{\perp}}\cdots
P_{R^{\perp}}A_{i_N}P_{R^{\perp}})
\end{eqnarray} since
$A_{i}P_{R^{\perp}}=P_RA_{i_N}P_{R^{\perp}}+P_{R^{\perp}}A_{i_N}P_{R^{\perp}}$
and the $P_R$ in the first summand goes through all the matrices
$A_{i_j}$ to finally cancel with $P_{R^{\perp}}$. Then we have
also matrices $C_i=P_{R^{\perp}}A_{i_N}P_{R^{\perp}}$ of size
$\dim(R^{\perp})\times \dim(R^{\perp})$ such that we can write out
original state with the following $D\times D$ matrices $$\left(%
\begin{array}{cc}
  B_i & 0  \\
  0 & C_i  \\
 \end{array}%
\right).$$

For each one of these blocks we reason similarly and we end up
with block-shaped matrices with the property that each block
satisfies $1$ in the Theorem. Let us now assume that for one of
the blocks, the map $X\mapsto \sum_i B_iXB_i^{\dagger}$ has a
fixed point $X\not =\mathbbm{1}$. We can suppose $X$ self-adjoint
and then diagonalize it $X=\sum_\alpha \lambda_\alpha
|\alpha\>\<\alpha|$ with $\lambda_1\ge\cdots \ge \lambda_n$.
Obviously, $\mathbbm{1}-\frac{1}{\lambda_1}X$ is a positive fixed
point that is not full rank, and this allows us (reasoning as
above) to decompose further the block $B_i$ in subblocks until
finally every block satisfies both properties 1 and 3 in the
Theorem. By the same arguments we can ensure that the only fixed
point of the dual map $X\mapsto \sum_i B_i^{\dagger}XB_i$ of each
block is also positive and full rank, and so, by choosing an
adequate unitary $U$ and changing $B_i$ to $UB_iU^{\dagger}$, we
can diagonalize this fixed point to make it a diagonal positive
full-rank matrix $\Lambda$, which finishes the proof of the
Theorem. }

\

Note that Thm.\ref{Th:TIcanonical} gives rise to a decomposition
of the state into a superposition of TI MPS each of which has only
one block in its canonical form and a respective cp-map $\E_j$
with a non-degenerate eigenvalue 1 (due to the uniqueness of the
fixed point). The following argument shows that in cases where
$\E_j$ has other eigenvalues of magnitude one further
decomposition into a superposition of periodic states is possible.

Examples of states with such periodic decompositions  (for $p=2$)
are the anti-ferromagnetic GHZ state and the Majumdar-Gosh state.

\begin{theorem}[Periodic decomposition]\label{Th:periodic}
Consider any TI state $\psi\in\mathbb{C}^{d\otimes N}$which has
only one block in its canonical TI MPS representation
(Thm.\ref{Th:TIcanonical}) with respective $D\times D$ matrices
$\{A_i\}$. If $\E(X)=\sum_iA_iXA_i^\dagger$ has $p$ eigenvalues of
modulus one, then if $p$ is a factor of $N$ the state can be
written as a superposition of $p$ $p$-periodic states each of
which has a MPS representation with bond dimension $D$. If $p$ is
no factor, then $\psi=0$.
\end{theorem}

\proof{ The theorem is a consequence of the spectral properties of
the cp map $\E$, which were proven in \cite{FaNaWe92}. There it is
shown that if the identity is the only fixed point of $\E$, then
there exists a $p\in\mathbb{N}$ such that $\{\omega^k\}_{k=1\ldots
p}$ with $\omega=\exp \frac{2\pi i}p$ are all eigenvalues of $\E$
with modulus 1. Moreover, there is a unitary $U=\sum_{k=1}^p
\omega^k P_k$, where $\{P_k\}$ is a set of orthogonal projectors
with $\sum_kP_k=\mathbbm{1}$ such that $\E(X P_k)=\E(X)P_{k-1}$
for all $D\times D$ matrixes $X$ (and cyclic index $k$). It is
straightforward to show that the latter implies that
\begin{equation}
\forall j,k:\ \ A_jP_k=P_{k-1}A_j\;.
\end{equation}
Exploiting this together with the decomposition of the trace
$\trace[\ldots]=\sum_k \trace[P_k\ldots P_k]$ leads to a
decomposition of the state $|\psi\rangle=\sum_{k=1}^p
|\psi_k\rangle$ where each of the states $|\psi_k\rangle$ in the
superposition has a MPS representation with site-dependent
matrices $A_{i_j}=P_{k+j-1}A_{i_j}P_{k+j}$. Hence, each
$|\psi_k\rangle$ is $p$-periodic and, since
$P_kP_l=\delta_{k,l}P_k$, non-zero only if $p$ is a factor of $N$.
}

\subsubsection{Generic cases}

Before proceeding we have to introduce two \emph{generic}
conditions on which many of the following results are based on.
The first condition is related to \emph{injectivity} of the map
\begin{equation}\label{eq:injectivity}
\Gamma_L:X\mapsto\sum_{i_1,\ldots,i_L=1}^d
\trace\big[XA_{i_1}\cdots A_{i_L}\big]\;|i_1\ldots i_L\rangle.
\end{equation}
Note that $\Gamma_L$ is injective iff the set of matrices
$\{A_{i_1}\cdots A_{i_L}: 1 \le i_1,\ldots,i_L\le d\}$ spans the
entire space of $D\times D$ matrices. Moreover, if $\sum_i
A_iA_i^\dagger=\mathbbm{1}$ then evidently injectivity of
$\Gamma_L$ implies injectivity of $\Gamma_{L'}$ for all $L'\geq
L$. To see the relation to `generic' cases consider $d$ randomly
chosen matrices $A_i$. The dimension of the span of their products
$A_{i_1}\cdots A_{i_L}$ is expected to grow as $d^L$ up to the
point where it reaches $D^2$. That is, for generic cases we expect
to have injectivity for $L\geq\frac{2\ln D}{\ln d}$. This
intuition can easily be verified numerically and rigorously proven
at least for $d=D=2$. However, in order not to rely on the vague
notion of `generic' cases we
introduce the following:\vspace{3pt}\\
{\bf Condition C1:} \emph{There is a finite number
$L_0\in\mathbb{N}$ such that $\Gamma_{L_0}$ is injective.\vspace{3pt}}\\
We continue by deriving some of the implications of condition C1
on the TI canonical form:
\begin{prop}\label{prop-inj} Consider a TI state represented in canonical MPS form (Thm.\ref{Th:TIcanonical}). If condition C1 is satisfied for $L_0<N$, then
\begin{enumerate}
\item we have only one block in the canonical representation.
\item if we divide the chain in two blocks of consecutive spins
$[1\ldots R],[R+1\ldots N]$, both of them with at least $L_0$
spins, then the rank of the reduced density operator
$\rho_{[1..R]}$ is exactly $D^2$.
\end{enumerate}
\end{prop}

\proof{ The first assert is evident, since any $X$ which has only
entries in the off-diagonal blocks would lead to
$\Gamma_{L_0}(X)=0$. To see the other implication we take our
$D$-MPS
\begin{equation*} |\psi\rangle=\sum_{i_1,\ldots,i_N}
\trace(A_{i_1}\cdots A_{i_N})|i_1\cdots i_N\rangle,
\end{equation*}
and introduce a resolution of the identity
$$\sum_{\alpha,\beta=1}^D\sum_{i_1,\ldots,i_N} \<\alpha|A_{i_1}\cdots
A_{i_R}|\beta\>\<\beta|A_{i_{R+1}}\cdots
A_{i_N}|\alpha\>|i_1\cdots i_N\rangle=\vspace*{-20pt}$$
$$=\sum_{\alpha,\beta=1}^D |\Phi_{\alpha,\beta}\>
|\Psi_{\alpha,\beta}\>,\quad\text{where}\vspace*{-10pt}$$
\begin{eqnarray}\nonumber
|\Phi_{\alpha,\beta}\>&=&\sum_{i_1,\ldots,i_R}
\<\alpha|A_{i_1}\cdots A_{i_R}|\beta\>|i_1\cdots i_R\rangle,\\
|\Psi_{\alpha,\beta}\>&=&\sum_{i_1,\ldots,i_R}
\<\beta|A_{i_R}\cdots A_{i_N}|\alpha\>|i_{R+1}1\cdots
i_N\rangle.\nonumber
\end{eqnarray}
It is then sufficient to prove that both
$\{|\Phi_{\alpha,\beta}\>\}$ and $\{|\Psi_{\alpha,\beta}\>\}$ are
sets of linearly independent vectors. But this is a consequence of
C1: Let us take complex numbers $c_{\alpha,\beta}$ such that
$\sum_{\alpha,\beta} c_{\alpha,\beta} |\Phi_{\alpha,\beta}\>=0$
(the same reasoning for the $|\Psi_{\alpha,\beta}\>$). This is
exactly
$$\Gamma_R\left[\sum_{\alpha,\beta}c_{\alpha,\beta}|\beta\>\<\alpha|\right]=0.$$  By C1 we have
that $\sum_{\alpha,\beta}c_{\alpha,\beta}|\beta\>\<\alpha|=0$ and
hence $c_{\alpha,\beta}=0$ for every $\alpha,\beta$. }

\

Now we will introduce a second condition for which we assume w.l.o.g. the spectral radius of $\E$ to be one:\vspace{3pt}\\
{\bf Condition C2:} \emph{The map $\E$ has only one eigenvalue of magnitude one.\vspace{3pt}}\\
 Again this is satisfied for `generic' cases as the set of cp maps with eigenvalues which are degenerated in magnitude is certainly of measure zero.
  It is shown
in \cite{FaNaWe92} that this condition is {\it essentially}
equivalent (Appendix \ref{open.problems}) to condition C1. In
particular C2 also implies that there is just one block in the TI
canonical representation (Thm.\ref{Th:TIcanonical}) of the MPS
$|\psi\>=\sum_{i_1,\ldots,i_N} \trace(A_{i_1}\cdots
A_{i_N})|i_1\cdots i_N\rangle$.

 Moreover, condition C2 implies that,
for sufficiently large $N$, we can approximate $E_{\mathbbm{1}}^N$
(which corresponds to $\E^N$ via Eq.(\ref{E2E})) by $|R\>\<L|$;
where $|R\>$ corresponds to the fixed point of $\E$ (that is, the
identity), and $\<L|$ to the fixed point $\Lambda$ of the dual
map.

Introducing a resolution of the identity as above, we have that
$|\psi\>=\sum_{\alpha,\beta}|\Psi_{\alpha,\beta}\>|\Psi_{\beta,\alpha}\>$,
with $|\Psi_{\alpha,\beta}\>=\sum_{i_1,\ldots,i_{\frac{N}{2}}}
\<\alpha|A_{i_1}\cdots A_{i_{\frac{N}{2}}}|\beta\>|i_1\cdots
i_{\frac{N}{2}}\rangle$. But now
$$\<\Psi_{\alpha,\beta}|\Psi_{\alpha',\beta'}\>=
\<\alpha'|\E^{\frac{N}{2}}(|\beta'\>\<\beta|)|\alpha\>=\lambda_{\alpha}\delta_{\alpha,\alpha'}\delta_{\beta,\beta'},$$
up to corrections of the order $|\nu_2|^{N/2}$ (where $\nu_2$ is
the second largest eigenvalue of $\E$). This implies that with
increasing $N$
$$|\psi\>=\sum_{\alpha,\beta}\sqrt{\lambda_\alpha \lambda_\beta}
\frac{|\Psi_{\alpha,\beta}\>}{\sqrt{\lambda_\alpha}}\frac{|\Psi_{\beta,\alpha}\>}{\sqrt{\lambda_\beta}}$$
becomes the Schmidt decomposition associated to half of the chain.
Hence we have proved the following.

\begin{theorem}[Interpretation of $\Lambda$]
Consider a TI MPS state. In the generic case (condition C2), the
eigenvalues of its reduced density operator with respect to half
of the chain converge with increasing $N$ to the diagonal matrix
$\Lambda\otimes \Lambda$ with $\Lambda$ from the TI canonical form
(Thm.\ref{Th:TIcanonical}).
\end{theorem}

\subsubsection{Uniqueness}

We will prove in this section that the TI canonical form in
Thm.\ref{Th:TIcanonical} is unique in the generic case.

\begin{theorem}[Uniqueness of the canonical form]\label{thm-uniq}

Let $$|\psi\>=\sum_{i_1,\ldots,i_N=0}^{d-1} \trace(B_{i_1}\cdots
B_{i_N})|i_1\cdots i_N\>$$ be a TI canonical $D$-MPS such that (i)
condition C1 holds, (ii) the OBC canonical representation of
$|\psi\>$ is unique, and (iii) $N>2L_0+D^4$ (a condition
polynomial in $D$). Then, if $|\psi\>$ admits another TI canonical
$D$-MPS representation
$$|\psi\>=\sum_{i_1,\ldots,i_N=0}^{d-1} \trace(C_{i_1}\cdots
C_{i_N})|i_1\cdots i_N\>,$$ there exists a unitary matrix $U$ such
that $B_i=UC_iU^{\dagger}$ for every $i$ (which implies in the
case where $\Lambda$ is non-degenerate that $B_i=C_i$ up to
permutations and phases).
\end{theorem}

To prove it we need a pair of lemmas.

\begin{lemma}\label{lem-same-matr}
Let $T,S$ be linear maps defined on the same vector spaces and
suppose that there exist vectors $Y_1,\ldots,Y_n$ such that
\begin{itemize}
\item $T(Y_k)=S(Y_{k+1})$ for every $1\le k\le n-1$, \item
$Y_1,\ldots,Y_{n-1}$ are linearly independent, \item
$Y_n=\sum_{k=1}^{n-1}\lambda_k Y_k$.
\end{itemize}
Consider a solution $x\neq 0$ of the equation $\lambda_1 x^{n-1}
+\cdots + \lambda_{n-1}x=1,$  and  define
\begin{align*}
\mu_1&=\lambda_1x\\
\mu_2&=\lambda_1x^2+\lambda_2x\\
\cdots\\
\mu_{n-1}&=\lambda_1 x^{n-1} +\cdots + \lambda_{n-1}x=1.
\end{align*}
Then, if $Y=\sum_{k=1}^{n-1}\mu_k Y_k$, we have that $Y\not =0$
and  $T(Y)=\frac{1}{x}S(Y)$.
\end{lemma}

\proof{Clearly
$$T(Y)=\sum_{k=1}^{n-1}\mu_kT(Y_k)=\sum_{k=1}^{n-1}\mu_kS(Y_{k+1})=$$
$$=S\left(\sum_{k=1}^{n-2}\mu_kY_{k+1}+\sum_{k=1}^{n-1}\lambda_kY_{k}\right)=$$
$$=S\left(\lambda_1Y_1+ (\lambda_2+\mu_1)Y_2+\cdots
+(\lambda_{n-1}+\mu_{n-2})Y_{n-1}\right),$$ and this last
expression is exactly $\frac{1}{x}S(Y)$ by the definition of the
$\mu$'s. Moreover, since $\mu_{n-1}=1$ and $Y_1,\ldots,Y_{n-1}$
are linearly independent we have that $Y\not = 0$. }

The following lemma is a consequence of \cite[Theorem
4.4.14]{HornJohnson91}.
\begin{lemma}\label{lem-horn}
If $B,C$ are square matrices of the same size $n\times n$, the
space of solutions of the matrix equation $$W(C\otimes
\mathbbm{1})=(B\otimes \mathbbm{1})W$$ is $S\otimes M_n$, where
$S$ is the space of solutions of the equation $XC=BX$.
\end{lemma}
We can prove now Theorem \ref{thm-uniq}.

\proof{ By Proposition \ref{prop-inj} we know that the matrices
$A^{[j]}_i$ in the canonical OBC representation of $|\psi\>$ are
of dimension $D^2\times D^2$ for any $L_0\le j\le N-L_0$ (in
particular there are at least $D^4$ of such $j$'s). From the TI
MPS representation of $|\psi\>$ we can obtain an alternative OBC
representation by noticing that
$$|\psi\>=\sum_{i_1,\ldots,i_N=0}^{d-1} b^{[1]}_{i_1}(B_{i_2}\otimes \mathbbm{1})\cdots
(B_{i_N}\otimes \mathbbm{1})b^{[N]}_{i_N}|i_1\cdots i_N\>,$$ where
$b^{[1]}_i$ is the vector $1\times D^2$ that contains all the rows
of $B_i$, that is, $$b^{[1]}_i
=(B_i(1,1),B_i(1,2),\ldots,B_i(1,D), B_i(2,1),\ldots),$$ and
$b^{[N]}_i$ is the vector $D^2\times 1$ that contains all the
columns of $B_i$, that is, $$(b^{[N]}_i)^T=
(B_i(1,1),B_i(2,1),\ldots,B_i(D,1), B_i(1,2),\ldots).$$

Doing the same with the $C$'s we have also
$$|\psi\>=\sum_{i_1,\ldots,i_N=0}^{d-1} c^{[1]}_{i_1}(C_{i_2}\otimes \mathbbm{1})\cdots
(C_{i_N}\otimes \mathbbm{1})c^{[N]}_{i_N}|i_1\cdots i_N\>.$$

Using now Theorem \ref{free-OBC} and the fact that between $L_0$
and $N-L_0$ both $A^{[j]}_i$, $(B_i\otimes \mathbbm{1})$ and
$(C_i\otimes \mathbbm{1})$ are $D^2\times D^2$ matrices, we can
conclude that there exist invertible $D^2\times D^2$ matrices
$W_1\ldots W_{D^4}$ such that $W_k(C_i\otimes
\mathbbm{1})=(B_i\otimes \mathbbm{1})W_{k+1}$ for every $1\le k\le
D^4-1$.

Now take $n$ such that $W_1,\ldots, W_{n-1}$ are linearly
independent but $W_n=\sum_{k=1}^{n-1}\lambda_k W_k$. Let us define
$x$ and $\mu_1,\ldots,\mu_{n-1}$ as in Lemma \ref{lem-same-matr}
and $W=\sum_{k=1}^{n-1}\mu_k W_k$. By Lemma \ref{lem-same-matr} we
have $W\not =0$ and $W(C_i\otimes
\mathbbm{1})=(\frac{1}{x}B_i\otimes \mathbbm{1})W$ for every $i$.
Now, Lemma \ref{lem-horn} implies that there exist $R\not =0$ such
that $RC_i=\frac{1}{x}B_iR$ for every $i$.

We can use now that $\Lambda=\sum_i B_i^{\dagger}\Lambda B_i$ to
prove that $\sum_iC_i^{\dagger}R^{\dagger}\Lambda
RC_i=\frac{1}{|x|^2}R^{\dagger}\Lambda R$. Since the completely
positive map $X\mapsto \sum_iC_i^{\dagger}XC_i$ is trace
preserving (and $R^{\dagger}\Lambda R\not =0$) one has that
$|x|^2= 1$.

Now, from $\mathbbm{1}=\sum_i C_iC_i^{\dagger}$, we obtain that
$\sum_{i}B_iRR^{\dagger}B_i^{\dagger}=RR^{\dagger}$. Since the
$B_i$'s have only one box (Proposition \ref{prop-inj}) we conclude
that $RR^{\dagger}=\mathbbm{1}$ so that $R$ is a unitary. }

\subsubsection{Obtaining the canonical form}

In the previous section we have implicitly used the ``freedom"
that one has in the choice of the matrices in the generic case. In
this section we will make this explicit (answering question
\ref{quest.can.1}) and will use it to show how to obtain
efficiently the canonical form (answering question
\ref{quest.can.3}).

Let us take a TI state $|\psi\>\in\mathbb{C}^{d\otimes N}$ such
that the rank of all the reduced density operators is bounded by
$D^2$. Clearly it can be stored using a MPS with OBC in $Nd$
$D^2\times D^2$ matrices. If we are in the generic case and this
state has a canonical form verifying condition C1, it would be
very convenient to have a way of obtaining it, since it allows us
to store the state using only $d$ $D\times D$ matrices!

In this section we will show how the techniques developed so far
allow us to do it by solving an independent of $N$ system of
$O(D^8)$ quadratic equations with $O(D^4)$ unknowns.

We will assume that the problem has a solution, that is, the state
has a TI canonical form with condition C1. We will also assume
that we are in the generic case in the sense that the OBC
canonical form is unique (no degeneracy in the Schmidt
Decomposition). Then, the algorithm to find it reads as follows:

We start with the $D^2\times D^2$ matrices
$A^{[L_0+1]}_i,\ldots,A^{[L_0+D^4]}_i$ of the OBC canonical form.

We solve the following system (S) of quadratic equations in the
unknows $Y_j, Z_{j+1}, B_i$, $j=L_0+1,\ldots,L_0+D^4$,
$i=1,\ldots,d$ ($Y_j, Z_j$  are $D^2\times D^2$ matrices and $B_i$
 $D\times D$ matrices).

\begin{align*} B_i\otimes \mathbbm{1}&= Y_j A_j^{[j]}
Z_{j+1}
&\forall \, i,j\\
Y_jZ_j&=\mathbbm{1} &\forall \, j\\
\sum_i B_iB_i^{\dagger}&=\mathbbm{1}
\end{align*}
and we have the following.
\begin{theorem}[Obtaining the TI canonical form] Consider any TI state
$|\psi\>$ with unique OBC canonical form and such that the rank of
each reduced density operator (of successive spins) is bounded by
$D^2$.
\begin{enumerate}
\item If there is a TI MPS representation verifying condition C1,
then the above systems (S) of quadratic equations has a solution.
\item Any solution of (S) gives us a TI $D$-MPS representation of
$|\psi\>$, that is related to \textit{the} canonical one by
unitaries ($A_i=UB_iU^{\dagger}$).
\end{enumerate}
\end{theorem}

\proof{ We have seen in the previous Section that the canonical
representation is a solution for (S). Now, if $B_i$'s are the
solution of (S) and $A_i$'s are the matrices of the canonical
representation, we have, reasoning as in the proof of Theorem
\ref{thm-uniq}, that there exists an $R\not = 0$ and an $x\not =0$
such that $RB_i=\frac{1}{x}A_iR$ for every $i$. Using that
$\sum_iB_iB_i^{\dagger}=\mathbbm{1}$, that $\Lambda=\sum_i
A_i^{\dagger}\Lambda A_i$ and that $\mathbbm{1}$ is the only fixed
point of $X\mapsto \sum_iA_iXA_i^{\dagger}$ we can conclude, as in
the proof of Theorem \ref{thm-uniq}, that $R$ is unitary. }

\section{Parent Hamiltonians}\label{Sec:Hamiltonians}

This section pretends to extend the results of the seminal paper
\cite{FaNaWe92} to the case of a finite chain. That is, we will
study when a certain MPS is the \textit{unique} ground state of
certain \textit{gapped} local hamiltonian.  However, since we deal
with a finite chain, the arguments given in \cite{FaNaWe92} for
the ``uniqueness" part are  no longer valid, and we have to find a
different approach. As in the previous section we will start with
the case of OBC and then move to the case of TI and PBC. In the
``gap" part we will simply sketch the original proof given in
\cite{FaNaWe92}.
\subsection{Uniqueness}\label{sec:Huniqueness}

\subsubsection{Uniqueness of the ground state under condition C1 in the case of OBC}

Let us take a MPS with OBC given in the canonical form
$|\psi\rangle=\sum_{i_1,\ldots,i_N} A^{[1]}_{i_1}\cdots
A^{[N]}_{i_N}|i_1\cdots i_N\rangle$. Let us assume that we can
group the spins in blocks of consecutive ones in such a way that,
in the regrouped MPS $|\psi\rangle=\sum_{i_1,\ldots,i_M}
B^{[1]}_{i_1}\cdots B^{[M]}_{i_M}|i_1\cdots i_M\rangle$, every set
of matrices $B^{[j]}_{i_j}$ verifies condition C1, that is,
generates the corresponding space of matrices. If we call
$h_{j,j+1}$ the projector onto the orthogonal subspace of
$$\{\sum_{i_j,i_{j+1}}\trace(XB^{[j]}B^{[j+1]}): X \text{ arbitrary}\},$$
then

\begin{theorem}[Uniqueness with OBC]
$|\psi\rangle$ is the unique ground state of the local Hamiltonian
$H=\sum_j h_{j,j+1}$.
\end{theorem}

\proof{ Any ground state $|\phi\>$ of $H$ verifies that
$h_{j,j+1}\otimes \1|\phi\>=0$ for every $j$, that is
\begin{equation}\label{thm:unique-OBC}|\phi\>= \sum_{i_1,\ldots,i_M}
\trace(X^j_{I(j,j+1)}B^{[j]}_{i_j}B^{[j+1]}_{i_{j+1}})|i_1\ldots
i_M\>,\end{equation} where $I(j,j+1)$ is the set of indices
$i_1\ldots i_{j-1}, i_{j+2} \ldots i_M$.

Mixing (\ref{thm:unique-OBC}) for $j$ and $j+1$ and using
condition C1 for $B^{[j+1]}$ gives $$X^j_{I(j,j+1)}B^{[j]}_{i_j}=
B^{[j+2]}_{i_{j+2}} X^{j+1}_{I(j+1,j+2)}.$$

Using now that $\sum_{i_j} B^{[j]}_{i_j} B^{[j]\,
\dagger}_{i_j}=\1$ and calling
$Y^j_{I(j,\ldots,j+2)}=\sum_{i_j}X^{j+1}_{I(j+1,j+2)} B^{[j]\,
\dagger}_{i_j}$ we get

\begin{equation*}|\phi\>= \sum_{i_1,\ldots,i_M}
\trace(Y^j_{I(j,\ldots,j+2)}B^{[j]}_{i_j}\cdots
B^{[j+2]}_{i_{j+2}})|i_1\ldots i_M\>.\end{equation*}

Using the trivial fact that blocking again preserves condition C1
one can easily finish the argument by induction. We just notice
that in the last step one obtains $$|\phi\>=\sum_{i_1,\ldots,i_M}
XB^{[1]}_{i_1}\cdots B^{[M]}_{i_M}|i_1\cdots i_M\rangle,$$ where
$X$ is just a number that, by normalization, has to be $1$, giving
$|\phi\>=|\psi\>$ and hence the result. }

\subsubsection{Uniqueness of the ground state under condition
C1 with TI and PBC}

To obtain the analogue result in the case of TI and PBC one can
apply the same argument. However, since we do not have any more
vectors in the first and last positions, we need to refine the
reasoning of the last step. Moreover, using the symmetry we have
now, one can decrease a bit the interaction length of the
Hamiltonian, from $2L_0$ to $L_0+1$.

\

Let us be a bit more concrete. Given our ring of $N$
$d$-dimensional quantum systems, $L\le N$ and a subspace $S$ of
${\C^d}^{\otimes L}$, we denote $H_S=\sum_{i=1}^N\tau^i(h_S)$,
where $h_S$ is the projection onto $S^{\perp}$ \footnote{For all
the reasonings it is enough to consider $\tilde{h}:{\C^d}^{\otimes
L}\lra {\C^d}^{\otimes L}$ positive such that $\ker{\tilde{h}}=S$.
We take $\tilde{h}=P_{S^{\perp}}$ for simplicity.}. If we start
with a MPS $|\psi\>=\sum_{i_1,\ldots,i_N}\trace(A_{i_1}\cdots
A_{i_N})|i_1\cdots i_N\>$ with property C1, we will consider
$L>L_0$ and, as before,  the subspace $\G_L^A$ (or simply $\G_L$)
formed by the elements $\sum_{i_1,\ldots,i_L}\trace(XA_{i_1}\cdots
A_{i_L})|i_1\cdots i_L\>$. It is clear that $H_{\G_L}|\psi\>=0$
and that $H_{\G_L}$ is frustration free. Moreover, if $N\ge 2L_0$
and $L>L_0$, then

\begin{theorem}[Uniqueness with TI and PBC]\label{uniqueGS}
$|\psi\>$ is the only ground state of $H_{\G_L}$.
\end{theorem}

\proof{ Reasoning as in the case of OBC one can easily see that
any ground state $|\phi\>$ of $H_{\G_L}$ is in ${\G}_N$, that is,
has the form $|\phi\>=\sum_{i_1,\ldots,i_{N}}\trace(XA_{i_1}\cdots
A_{i_{N}})|i_1\cdots i_{N}\>$. Since there is no distinguished
first position, $|\phi\>$ can also be written
$$|\phi\>=\sum_{i_1,\ldots,i_{N}}\trace(A_{i_1}\cdots
A_{i_{L_0}}YA_{i_{L_0+1}}\cdots A_{i_{N}})|i_1\cdots i_{N}\>.$$ By
condition C1, $XA_{i_1}\cdots A_{i_{L_0}}=A_{i_1}\cdots
A_{i_{L_0}}Y$ for every $i_1,\ldots,i_{L_0}$. But, also by
condition C1, $A_{i_1}\cdots A_{i_{L_0}}$ generates the whole
space of $D\times D$ matrices. Hence $X=Y$ which, in addition,
commutes with $A_{i_1}\cdots A_{i_{L_0}}$ for every
$i_1,\ldots,i_{L_0}$. This means that $X$ commutes with every
matrix and hence $X=\lambda\mathbbm{1}$ and $|\phi\>=|\psi\>$. }

\subsubsection{Non-uniqueness in the case of two or more blocks}

In the absence of condition C1 in our MPS, we cannot guarantee
uniqueness for the ground state of the parent Hamiltonian. There
are two different properties that can lead to degeneracy. One is
the existence of a periodic decomposition (Theorem
\ref{Th:periodic}), that can happen even in the case of one block.
This is the case of the Majumdar-Gosh model (Section
\ref{Sec:prelim}). The other property is the existence of more
than one block in the canonical form (Thm.\ref{Th:TIcanonical}).
As we will see below, this leads to a stronger version of
degeneracy that is closely related to the number of blocks. In
particular, we are going to show that whenever we have more than
one block, the MPS is \textit{never} the unique ground state of a
frustration free local Hamiltonian (Thm \ref{2blocks.1}). In
addition, there exists one local Hamiltonian which has the MPS as
ground state and with ground space degeneracy equal to the number
of blocks (Thm \ref{2blocks.2}). That is, a number of blocks
greater than one can correspond to a spontaneously broken symmetry
\cite{NachtergaeleSym}.

\

In all this section we need to assume that we have condition C1 in
each block. For a detailed discussion of the
\textit{reasonability} of this hypothesis see Appendix
\ref{open.problems}.

\

Let us take a TI MPS $|\phi\>$ with $b (\ge 2)$ blocks
$A^1_i,\ldots, A_i^b$ in the canonical form, $\size(A^1_i)\ge
\cdots\ge \size(A^b_i)$, and condition C1 in each block. Let us
call $L_0=\max_j\{L_0^{A^j}\}$, which will be logarithmic in $N$
for the \textit{generic} case. Clearly
$|\phi\>=\sum_{j=1}^b|\phi_{A^j}\>$, where
$$|\phi_{A^j}\>=\lambda_j^N\sum_{i_1,\ldots,i_N}
\trace(A^j_{i_1}\cdots A^j_{i_N})|i_1\cdots i_N\rangle.$$
Moreover, w.l.o.g. we can assume that the states $|\phi_{A^j}\>$
are pairwise different. The following lemmas will take care of the
technical part of the section.

\begin{lemma}\label{lem1}
Given any $D\times D$ matrices $C\not =0$ and $X$ there exist
matrices $R_i, S_i$ such that $X=\sum_i R_iCS_i$.
\end{lemma}

\proof{ By the polar decomposition, it is easy to find matrices
$E,F,G, H$ (with $G,H$ invertible) such that $ECF=|1\>\<1|$ and
$GXH=\sum_{i=1}^m|i\>\<i|$. Clearly
$\sum_{i=1}^m|i\>\<i|=\sum_{i=1}^m P_i|1\>\<1|P_i,$ where $P_i$ is
the matrix obtained from the identity by permuting the first and
the $i$-th row. So $R_i=G^{-1}P_iE$ and $S_i=FP_iH^{-1}$. }

\begin{lemma}\label{lem:direct-sum}
If $L\ge 3(b-1)(L_0+1)$, the sum $\bigoplus_{j=1}^b
\mathcal{G}_L^{A^j}$ is direct.
\end{lemma}

\proof{ We group the spins in $3(b-1)$ blocks of at least $L_0+1$
spins each and then use induction. First the case $b=2$.

\

We assume on the contrary that there exist $X, Y\not =0$ such that
$$\sum_{i_{1},i_2,i_3}
\trace(A^1_{i_1}XA^1_{i_2}A^1_{i_3})|i_{1}i_2
i_3\rangle=\sum_{i_{1},i_2,i_3}
\trace(A^2_{i_1}YA^2_{i_2}A^2_{i_3})|i_{1}i_2 i_3\rangle.$$ If we
consider now an arbitrary matrix $Z$, by C1 and Lemma \ref{lem1},
there exist complex numbers $\mu^i_{i_1}, \rho^i_{i_2}$ such that
$Z=\sum_i\sum_{i_1,i_2}\mu^i_{i_1}
\rho^i_{i_{2}}A^1_{i_1}XA^1_{i_{2}}.$ Calling
$W=\sum_i\sum_{i_1,i_2}\mu^i_{i_1}
\rho^i_{i_{2}}A^2_{i_1}YA^2_{i_{2}}$ we have that
$\sum_{i_3}\trace(ZA^1_{i_3})|i_3\>=\sum_{i_3}\trace(WA^2_{i_3})|i_3\>$.

This means that $\G^{A^1}_{L_0+1}\subset\G^{A^2}_{L_0+1}$. Since
$\size(A^1_i)\ge \size(A^2_i)$, this implies that $\size(A^1_i)=
\size(A^2_i)$ and that $\G^{A^1}_{L_0+1}=\G^{A^2}_{L_0+1}$. But
now, taking the local Hamiltonian
$H_{\G_{L_0+1}^{A^1}}=H_{\G_{L_0+1}^{A^2}}$, by Theorem
\ref{uniqueGS}, both $|\phi_{A^1}\>$ and $|\phi_{A^2}\>$ should be
its \textit{only} ground state; which is the desired
contradiction.

\

Now the induction step. Let us start with $\sum_{j=1}^{b+1}w^j=0$,
where
$$\G^{A^j}_{3b(L_0+1)}\ni
w_j=\sum_{i_1,\ldots,i_{3b}}\trace(A^j_{i_1}W^jA^j_{i_2}\cdots
A^j_{i_{3b}})|i_1\cdots i_{3b}\>.$$ We want to prove that $W^j=0$
for every $j$. So let us assume the opposite, take $j$ such that
$W^j\not =0$ and call $\tilde{w}_j=\left(\1_{[12]}\otimes
h_{\G^{A^{b+1}}_{[3]}}\otimes \1_{[4\ldots 3b]}\right)(w_j)$. We
have that $\sum_{j=1}^b\tilde{w}_j=0$, and, by the induction
hypothesis, each $\tilde{w}_j=0$. Now
$$\tilde{w}_j=\sum_{i_1,\ldots,i_{3b}}\trace(A^j_{i_1}W^jA^j_{i_2}X^j_{i_3}A^j_{i_4}\cdots
A^j_{i_{3b}})|i_1\cdots i_{3b}\>,$$ with $X^j_{i_3}\not = 0$ for
some $i_3$ (Theorem \ref{uniqueGS}). Moreover, we can use
condition C1 and Lemma \ref{lem1} to get complex numbers
$\mu^i_{i_1}, \rho^i_{i_2}$ such that
$\1=\sum_i\sum_{i_1,i_2}\mu^i_{i_1}
\rho^i_{i_{2}}A^j_{i_1}W^jA^j_{i_{2}}.$

Hence, $0=\sum_{i_4,\ldots, i_{3b}}\trace(X_{i_3}^jA^j_{i_4}\cdots
A^j_{i_{3b}})|i_4\cdots i_{3b}\>$ for every $i_3$, which implies
by C1 the contradiction $X^j_{i_3}=0$. }

Finally the results,

\begin{theorem}[Degeneracy of the ground space v1]\label{2blocks.1}
If $N\ge 3(b-1)(L_0+1)+L$ and $H=\sum_i\tau^i(h)$ is any
translationally invariant frustration free $L$-local Hamiltonian
on our ring of $N$ spins that has $|\phi\>$ as a ground state
(that is, $H|\phi\>=0$), then  $|\phi_{A^j}\>$ is also a ground
state of $H$ for every $j$.

In particular $H$ has more than one ground state.
\end{theorem}

\proof{ One has $0=(h\otimes \1)|\phi\>=\sum_j(h\otimes
\1)|\phi_{A^j}\>$. Since $(h\otimes \1)|\phi_{A^j}\> \in
\left(\C^d\right)^{\otimes L}\otimes \G^{A^j}_{N-L}$, we can use
Lemma \ref{lem:direct-sum} to get the desired conclusion:
$(h\otimes \1)|\phi_{A^j}\>=0$ for every $j$.}

\begin{theorem}[Degeneracy of the ground space v2]\label{2blocks.2}
There exists a local Hamiltonian $H$ acting on $L\ge
3(b-1)(L_0+1)+1$ spins such that its ground space is exactly
$\ker(H)=\spanned\{|\phi_{A^j}\>\}_{1\le j\le b}$.
\end{theorem}

\proof{ The hamiltonian will be $H_S$ with $S=\bigoplus_j
\mathcal{G}^{A^j}_{L}$, and $L\ge 3(b-1)(L_0+1)+1$. For $m\ge L$,
$$\C^d\otimes\left(\bigoplus_j \mathcal{G}^{A^j}_{m}\right)\cap
\left(\bigoplus_j \mathcal{G}^{A^j}_{m}\right)\otimes
\C^d=\bigoplus_j \mathcal{G}^{A^j}_{m+1}.$$

In fact, if $|\phi\>\in \C^d\otimes(\bigoplus_j
\mathcal{G}^{A^j}_{m})\cap (\bigoplus_j
\mathcal{G}^{A^j}_{m})\otimes \C^d$, we have simultaneously that
\begin{align*}
|\psi\>&=\sum_j
\sum_{i_1,\ldots,i_{m+1}}\trace(A^j_{i_{m+1}}C^j_{i_1}A^j_{i_2}\cdots
A^j_{i_m}) |i_1\cdots i_{m+1}\>\quad \text{ and}\\
|\psi\>&=\sum_j\sum_{i_1,\ldots,i_{m+1}}\trace(D^j_{i_{m+1}}A^j_{i_1}A^j_{i_2}\cdots
A^j_{i_m})|i_1\cdots i_{m+1}\>\;.
\end{align*}
Lemma \ref{lem:direct-sum} and condition C1 allows us to identify
for every $j, i_1, i_{m+1}$
$$A^j_{i_{m+1}}C^j_{i_1}=D^j_{i_{m+1}}A^j_{i_1}.$$
Calling $E^j=\sum_{i_1}C^j_{i_1}A^j_{i_1}$ and using that
$\sum_{i_1}A^j_{i_1}A^{j\, \dagger}_{i_1}=\1$, we get
$A^j_{i_{m+1}}C^j_{i_1}=A^j_{i_{m+1}}E^jA^j_{i_1},$ which implies
that $|\phi\>\in \bigoplus_j \mathcal{G}^{A^j}_{m+1}.$

Then one can easily follow the lines of the proof of Theorem
\ref{uniqueGS} (assuming $N\ge L+L_0$) to conclude that
$\ker(H_S)= \spanned\{|\phi_{A^j}\>\}_{1\le j\le b}$. }

\subsection{Energy gap}\label{Sec:Hgap}

If the ground state energy is zero (which can always be achieved
by a suitable offset),  the \emph{energy gap} $\gamma$ above the
ground space is the largest constant for which
\begin{equation} H^2\geq\gamma H\;.
\end{equation}

If in addition $H=\sum_i \tau^i(h)$ is frustration free and has
interaction length $l$, by taking any $p\ge l$ and grouping the
spins in blocks of $p$, one can define an associated $2$-local
interaction in the regrouped chain by
$\tilde{h}_{i,i+1}=H_{[pi+1,\ldots, (i+2)p]}$ (
$=\sum_{j=pi}^{p(i+2)-l}\tau^j(h)$). The {\it new} hamiltonian
$\tilde{H}$ verifies $$H_{[1,\ldots, mp]}\le \tilde{H}_{[1\ldots
m]}=\sum_{i=0}^{m-1}\tau^i(\tilde{h})\le 2H_{[1,\ldots, mp]}.
$$
Moreover, calling $P$ to the projection onto $\ker(\tilde{h})$,
there exists a constant $\gamma_{2p}$ (that is exactly the
spectral gap of $H_{[1\ldots,2p]}$) such that $\tilde{h}\ge
\gamma_{2p}(\1-P)$.

Therefore, to study the existence of an energy gap in a local TI
frustration free Hamiltonian, it is enough to study the case of a
nearest neighbor interaction $\hat{H}=\sum_i\tau^i(P_{i,i+1})$
where $P$ is a projector. In this situation, Knabe \cite{Knabe}
gave a sufficient condition to assure the existence of a gap,
namely that the gap $\epsilon_n$ of $\hat{H}_{[1\ldots, n+1]}$ is
bigger than $\frac{1}{n}$ for some $n$.

In the particular case of the parent hamiltonian of an MPS, a much
more refined argument was provided in \cite{FaNaWe92} to prove the
existence of a gap under condition C1. The idea reads as follows.
Clearly $\hat{H}^2\ge \hat{H}+ \sum_{i} P_{i,i+1}P_{i+1, i+2}+
P_{i,i+2}P_{i+1, i+1}$. By the proof of Theorem \ref{uniqueGS},
$$P_{i,i+1}=\1_{[1\ldots pi]}\otimes (\1-P_{\G_{2p}}) \otimes \1_{[p(i+2)+1\ldots N]}.$$
Then a technical argument proves that $P_{i,i+1}P_{i+1, i+2}+
P_{i,i+2}P_{i+1, i+1}\ge -O(\nu_2^p)( P_{i,i+1}+ P_{i+1, i+2})$,
where $\nu_2$ is the second largest eigenvalue of $\E$. This gives
$\hat{H}^2\ge(1-O(\nu_2^p))\hat{H}$, which concludes the argument.

\section{Generation of MPS}\label{sec:generation}

The MPS formalism is particularly suited for the description of
sequential schemes for the generation of multipartite states.
Consider for instance a chain of spins in a pure product state.
Two possible \emph{sequential} ways of preparing a more general
state on the spin chain are either to let an ancillary particle
(the head of a Turing machine) interact sequentially with all the
spins or to make them interact themselves in a sequential manner:
first spin 1 with 2 then 2 with 3 an so on.

Clearly, many physical setups for the generation of multipartite
states are of such sequential nature: time-bin photons leaking out
of an atom-cavity system, atoms passing a microwave cavity or
laser pulses propagating through atomic ensembles. We will see in
the following that the MPS formalism provides the natural language
for describing such schemes. This section reviews and extends the
results obtained in \cite{SSVCW}. A detailed application of the
formalism to particular physical systems can be found in
\cite{SSVCW,SSVCW2}.

\subsection{Sequential generation with ancilla}

Consider a spin chain which is initially in a product state
$|0\rangle^{\otimes N}\in {\cal H}_{\cal B}^{\otimes N}$ with
${\cal H}_{\cal B}\simeq\mathbb{C}^d$ and an additional ancillary
system in the state $ |\varphi_I\rangle\in{\cal H}_{\cal A}\simeq
\mathbb{C}^D$. Let $\sum_{i,\alpha,\beta} A^{[k]}_{i,\alpha,\beta}
|\alpha,i\rangle\langle\beta,0|$ be a general stochastic operation
on ${\cal H}_{\cal A}\otimes {\cal H}_{\cal B}$ applied to the
ancillary system and the $k$'th site of the chain. This operation
could for instance correspond to one branch of a measurement or to
a unitary interaction in which case $\sum_i
A^{[k]\dag}_iA^{[k]}_i=\mathbbm{1}$. If we let the ancillary
system interact sequentially with all $N$ sites and afterwards
measure the ancilla in the state $|\varphi_F\rangle$, then the
remaining state on ${\cal H}_{\cal B}^{\otimes N}$ is (up to
normalization) clearly given by the MPS
\begin{equation}\label{OBCMPSgen}
|\psi\rangle=\sum_{i_1,\ldots,i_N=1}^d
\langle\varphi_F|A^{[N]}_{i_N}\cdots
A^{[2]}_{i_{2}}A^{[1]}_{i_1}|\varphi_I\rangle|i_N\cdots
i_1\rangle.
\end{equation}
By imposing different constraints on the allowed operations (and
thus on the $A$'s) we can distinguish the following types of
sequential generation schemes for pure multipartite states:
\begin{enumerate}
    \item \emph{Probabilistic schemes}: arbitrary stochastic operations with a $D$-dimensional ancilla are allowed.
    \item \emph{Deterministic schemes}: the interactions must be unitary and
    the $D$-dimensional ancilla must decouple in the last step (without measurement).
    \item \emph{Deterministic transition schemes} (for $d=2$): Here we
    consider an enlarged ancillary system ${\cal H}_{\cal A}=\mathbb{C}^D\oplus\mathbb{C}^D\simeq
    \mathbb{C}^D\otimes\mathbb{C}^2$ (corresponding e.g. to $D$ `excited' and $D$ `ground state'
    levels) and a fixed interaction of the form
    \begin{eqnarray}
    |\varphi\rangle|1\rangle|0\rangle_{\cal B} &\mapsto& |\varphi\rangle|0\rangle|1\rangle_{\cal
    B}\nonumber\;,\\
|\varphi\rangle|0\rangle|0\rangle_{\cal B} &\mapsto&
|\varphi\rangle|0\rangle|0\rangle_{\cal
    B}\nonumber\;.
    \end{eqnarray}
Moreover, we allow for arbitrary unitaries on $\cal H_A$ in every
step and require the ancilla to decouple in the last step.
\end{enumerate}
\begin{theorem}[Sequential generation with ancilla]\label{Thm:seqwith} The three sets of
multipartite states which can be generated by the above sequential
schemes with ancilla are all equal to the set of states with OBC
MPS representation with maximal bond dimension $D$.
\end{theorem}
Note that the proof of this statement in \cite{SSVCW} (based on
subsequent singular value decompositions) also provides a
\emph{recipe} for the generation of any given state (with minimal
resources). This idea has been recently exploited in \cite{Kike}
to analyze the resources needed for sequential quantum cloning.

\subsection{Sequential generation without ancilla}\label{Sec:gen2}

Let us now consider sequential generation schemes without ancilla.
The initial state is again a product $|0\rangle^{\otimes N} \in
\mathbb{C}^{d\otimes N}$ and we perform first an operation
affecting the sites 1 and 2, then 2 and 3  up to one between $N-1$
and $N$. Again we may distinguish between probabilistic and
deterministic schemes and as before both classes coincide with a
certain set of MPS:
\begin{theorem}
[Sequential generation without ancilla] The sets of pure states
which can be generated by a sequential scheme without ancilla
either deterministically or probabilistically  are both equal to
the class of states having an OBC MPS representation with bond
dimension $D\leq d$.
\end{theorem}

\proof{ Let us denote the map acting on site $k$ and $k+1$ by
$U^{[k]}$. Then for $k<N-1$ we can straight forward identify the
matrices in the MPS representation by
$A^{[k]}_{i,\alpha,\beta}=\langle i, \beta
|U^{[k]}|\alpha,0\rangle$ where for $k=1$ we have $\alpha=0$,
i.e., $A^{[1]}_i$ are  vectors. From the last map with
coefficients $\langle i,j|U^{[N-1]}|\alpha,0\rangle$ we obtain the
$A$'s by a singular value decomposition (in $i|\alpha,j$), such
that $A_i^{[N]}$ is again a set of vectors. Hence, all states
generated in this way are OBC MPS with $D\leq d$.

Conversely, we can generate every such MPS deterministically in a
sequential manner without ancilla. To do this we exploit
Thm.\ref{Thm:seqwith} and use the site $k+1$ as `ancilla' for the
$k$'th step (i.e., the application of a unitary $U^{[k]}$)
followed by a swap between site $k+1$ and $k+2$.  $N-1$ of these
steps are sufficient since the last step in the proof for the
deterministic part in Thm.\ref{Thm:seqwith} is just a swap between
the ancilla and site $N$. }

\section{Classical simulation of quantum systems}\label{sec:classical-sim}

We saw in Section \ref{Sec:prelim} that if a quantum many-body
state has a MPS representation with sufficiently small bond
dimension $D$, then we can efficiently store it on a classical
computer and calculate expectation values. The practical relevance
of MPS representations in the context of classical simulation of
quantum systems stems then from two facts: (i) Many of the states
arising in
 condensed matter theory (of one-dimensional systems) or
quantum information theory either have such a small-$D$ MPS
representation or are well approximated by one; and (ii) One can
efficiently obtain such approximating MPS.

The following section will briefly review the most important
results obtained along these lines.

\subsection{Properties of ground states of spin chains}
The main motivation for introducing the class of MPS was to find a
class of wavefunctions that capture the physics needed to describe
the low-energy sector of local quantum Hamiltonians. Once such a
class is identified and expectation values of all states in the
class can be computed efficiently, it is possible to use the
corresponding states in a variational method. The very successful
renormalization group methods, first developed by Wilson
\cite{Wilson} and later refined by White \cite{White}, are
precisely such variational methods within the class of matrix
product states \cite{Ostlund,vonDelft,VPC04}.

In the case of 1-D systems (i.e. spin chains) with local
interactions, the low-energy states indeed exhibit some remarkable
properties. First, ground states have by definition extremal local
properties (as they minimize the energy), and hence their local
properties determine their global ones. Let us consider any local
Hamiltonian of $N$ spins that exhibits the property that there is
a unique ground state $|\psi_{ex}\rangle$ and that the  gap is
$\Delta(N)$. Let us furthermore consider the case when $\Delta(N)$
decays not faster than an inverse polynomial in $N$ (this
condition is satisfied for all gapped systems and for all known
critical translationally invariant systems). Then let us assume
that there exists a state $|\psi_{appr}\rangle$ that reproduces
well the local properties of all nearest neighbor reduced density
operators: $\|\rho_{appr}-\rho_{ex}\|\leq \delta$. Then it follows
that the global overlap is bounded by
\[\||\psi_{ex}\rangle-|\psi_{appr}\rangle\|^2\leq \frac{N\delta}{\Delta(N)}.\] This is remarkable as it shows that it is enough to
reproduce the local properties well to guarantee that also the
global properties are reproduced accurately: for a constant global
accuracy $\epsilon$, it is enough to reproduce the local
properties well to an accuracy $\delta$ that scales as an inverse
polynomial in the number of spins. This is very relevant in the
context of variational simulation methods: if the energy is well
reproduced and if the computational effort to get a better
accuracy in the energy only scales polynomially in the number of
spins, then a scalable numerical method can be constructed that
reproduces all global properties well (here scalable means
essentially a polynomial method) \footnote{Of course this does not
apply to global quantities, like entropy, where one needs
exponential accuracy in order to have closeness}.

Second, there is very few entanglement present in ground states of
spin chains, even in the case of a critical system. The relevant
quantity here is to study area-laws: if one considers the reduced
density operator $\rho_L$ of a contiguous block of $L$ spins in
the ground state of a spin chain with $N\gg L$ spins, how does the
entropy of that block scale with $L$? This question was first
studied in the context of black-hole entropy
\cite{Bombelli-Srednicki-Bousso} and has recently attracted a lot
of attention \cite{arealaw, Ortiz, Cardy}. Ground states of local
Hamiltonians of spins seem to have the property that the entropy
is not an extensive property but that the leading term in the
entropy only scales as the boundary of the block (hence the name
area-law), which means a constant in the case of a 1-D system
\cite{Cardy, Jin}:

\begin{equation} S^\alpha(\rho_L)\simeq
\frac{c}{6}\left(1+\frac{1}{\alpha}\right)\log(\xi)\label{arealaw}\end{equation}

Here $S_\alpha$ is the Renyi entropy
\[S^\alpha(\rho)=\frac{1}{1-\alpha}\log\left({\rm Tr}\rho^\alpha\right),\]
$c$ is the central charge\footnote{We note that Eq.(\ref{arealaw})
has been proven for critical spin chains (in particular the
XX-model with transverse magnetical field \cite{Jin}) which are
related to a conformal field theory. A general result is still
lacking.} and $\xi$ the correlation length. This has a very
interesting physical meaning: it shows that most of the
entanglement must be concentrated around the boundary, and
therefore there is much less entanglement than would be present in
a random quantum state (where the entropy would be extensive and
scale like $L$). The area law (\ref{arealaw}) is mildly violated
in the case of 1-D critical spin systems where $\xi$ has to be
replaced with $L$, but even in that case the amount of
entanglement is still exponentially smaller than the amount
present in a random state. This is very encouraging, as one may
exploit the lack of entanglement to simulate these systems
classically. Indeed, we already proved that MPS obey the same
property \cite{Ortiz}.

The existence of an area law for the scaling of entropy is
intimately connected to the fact that typical quantum spin systems
exhibit a finite correlation length. In fact, it has been recently
proven \cite{hastings1} that all connected correlation functions
between two blocks in a gapped system have to decay exponentially
as a function of the distance of the blocks. Let us therefore
consider a 1-D gapped quantum spin system with correlation length
$\xi_{corr}$. Due to the finite correlation length, it is expected
that the reduced density operator $\rho_{AB}$ obtained when
tracing out a block $C$ of length $l_{AB}\gg \xi_{corr}$ is equal
to
\begin{equation} \rho_{AB}\simeq
\rho_A\otimes\rho_B\label{hasg}\end{equation} up to exponentially
small corrections \footnote{Strictly speaking, Hastings theorem
does not imply the validity of equation (\ref{hasg}), as it was
shown in \cite{datahiding} that orthogonal states exist whose
correlation functions are exponentially close to each other;
although it would be very surprising that ground states would
exhibit that property, this prohibits to turn the present argument
into a rigorous one.}. The original ground state
$|\psi_{ABC}\rangle$ is a purification of this mixed state, but it
is of course also possible to find another purification of the
form $|\psi_{AC_l}\rangle\otimes|\psi_{BC_r}\rangle$ (up to
exponentially small corrections) with no correlations whatsoever
between $A$ and $B$; here $C_l$ and $C_r$ together span the
original block $C$. Since different purifications are unitarily
equivalent, there exists a unitary operation $U_C$ on the block
$C$ that completely disentangles the left from the right part:
\[I_A\otimes U_{C}\otimes I_C|\psi_{ABC}\rangle\simeq|\psi_{AC_l}\rangle\otimes|\psi_{BC_r}\rangle.\]
This implies that there exists a tensor $A_{i,\alpha,\beta}$ with
indices $1\leq i,\alpha,\beta\leq D$ (where $D$ is the dimension
of the Hilbert space of $C$) and states
$|\psi^A_\alpha\rangle,|\psi_i^C\rangle,|\psi^B_\beta\rangle$
defined on the Hilbert spaces belonging to $A,B,C$ such that
\[|\psi_{ABC}\rangle\simeq \sum_{i,\alpha,\beta}A_{i,\alpha,\beta}|\psi^A_\alpha\rangle|\psi^C_i\rangle|\psi^B_\beta\rangle.\]
Applying this argument recursively leads to a matrix product state
description of the state and gives a strong hint that ground
states of gapped Hamiltonians are well represented by MPS. It
turns out that this is even true for critical systems.

\subsection{MPS as a class of variational wavefunctions} \label{Sec:approx}

Here we will review the main results of \cite{VeCi05}, which give
analytical bounds for the approximation of a state by a MPS that
justify the choice of MPS as a reasonable class of variational
wavefunctions.

We consider an arbitrary state $|\psi\rangle$ and denote by
\[\{\mu^{[k]i}\}, i=1..N_k=d^{\min(k,N-k)}\] the eigenvalues of
the reduced density operators
\[\rho_k=\rm{Tr}_{k+1,k+2,\ldots
,N}|\psi\rangle\langle\psi|,\] sorted in decreasing order.

\begin{theorem}
There exists a MPS $|\psi_D\rangle$ with bond dimension $D$ such
that
 \[\||\psi\rangle-|\psi_D\rangle\|^2\leq
2\sum_{k=1}^{N-1}\epsilon_k(D)
\]
where $\epsilon_k(D)=\sum_{i=D+1}^{N_k}\mu^{[k]i}$.
\end{theorem}

This shows that for systems for which the $\epsilon_k(D)$ decay
fast in $D$, there exist MPS with \emph{small} $D$ which will not
only reproduce well the local correlations (such as energy) but
also all the nonlocal properties (such as correlation length).

The next result relates the derived bound to the Renyi entropies
of the reduced density operators. Given a density matrix $\rho$,
we denote as before $\epsilon(D)=\sum_{i=D+1}^\infty\lambda_i$
with $\lambda_i$ the nonincreasingly ordered eigenvalues of
$\rho$. Then we have

\begin{theorem}
If $0<\alpha<1$, then $\log(\epsilon(D))\leq
\frac{1-\alpha}{\alpha}\left(S^\alpha(\rho)-\log\frac{D}{1-\alpha}\right)$.
\end{theorem}

The two results together allow us to investigate the computational
effort needed to represent critical  systems, arguable the hardest
ones to simulate \footnote{For non--critical systems, the
renormalization group flow is expected to increase the Renyi
entropies in the UV direction. The corresponding fixed point
corresponds to a critical system whose entropy thus upper bounds
that of the non--critical one.}, as MPS. The key fact here is the
area-law (\ref{arealaw}), which for critical systems reads
 \begin{equation}
 S^\alpha(\rho_L)\simeq\frac{c}{6}
 \left(1+\frac{1}{\alpha}\right)\log(L)\label{infg}
 \end{equation}
for all $\alpha>0$ ($c$ the central charge).

Let us therefore consider the Hamiltonian associated to a critical
system, but restricted to $2L$ sites. The entropy of a half chain
(we consider the ground state $|\psi_{ex}\rangle$ of the finite
system) will typically scale as in eq. (\ref{infg}) but with an
extra term that scales like $1/L$. Suppose we want to get that
$\||\psi_{ex}\rangle-|\psi_D\rangle\|^2\leq\epsilon_0/L$ with
$\epsilon_0$ independent of $L$ \footnote{We choose the $1/L$
dependence such as to assure that the absolute error in extensive
observables does not grow.}. If we call $D_L$ the minimal $D$
needed to obtain this precision for a chain of length $2L$, the
previous two results combined yield
\begin{theorem}
\[D_L\leq
cst\left(\frac{L^2}{(1-\alpha)\epsilon_0}\right)^{\frac{\alpha}{1-\alpha}}
L^{\frac{c+\bar{c}}{12}\frac{1+\alpha}{\alpha}}.\]
\end{theorem}
This shows that $D$ only has to scale polynomially in $L$ to keep
the accuracy $\epsilon_0/L$ fixed; in other words, there exists an
efficient scalable representation for ground states of critical
systems (and hence also of noncritical systems) in terms of MPS.
This is a very strong result, as it shows that one can represent
ground states of spin chains with only polynomial effort (as
opposed to the exponential effort if one would do e.g. exact
diagonalization).

The above result was derived from a certain scaling of the block
Renyi entropy with $\alpha<1$. In fact, for $\alpha\geq 1$, i.e.,
in particular for the von Neumann entropy ($\alpha=1$), even a
saturating block entropy does in general not allow to conclude
that states are efficiently approximable by MPS \cite{entropyMPS}.
Conversely, a linear scaling for the von Neumann entropy, as
generated by particular time evolutions, rules out such an
approximability \cite{entropyMPS}.

\subsection{Variational algorithms}
Numerical renormalization group methods have since long been known
to be able to simulate spin systems, but it is only recently that
the underlying structure of matrix product states has been
exploited. Both NRG, developed by K. Wilson \cite{Wilson} in the
'70s, and DMRG, developed by S. White \cite{White} in the '90s,
can indeed be reformulated as variational methods within the class
of MPS \cite{VPC04,vonDelft}. The main question is how to find the
MPS that minimizes the energy for a given spin chain Hamiltonian:
given a Hamiltonian $\mathcal{H}$ acting on nearest neighbours, we
want to find the MPS $|\psi\rangle$ such that the energy
\[\frac{\langle\psi|\mathcal{H}|\psi\rangle}{\langle\psi|\psi\rangle}\]
is minimized. If $|\psi\rangle$ is a TI MPS, then this is a highly
complex optimization problem. The main trick to turn this problem
into a tractable one is to break the translational symmetry and
having site-dependent matrices $A_i^{[k]}$ for the different spins
$k$; indeed, then the functional to be minimized is a
multiquadratic function of all variables, and then one can use the
standard technique of alternating least squares to do the
minimization \cite{VPC04}. This works both in the case of open and
closed boundary conditions, and in practice the convergence of the
method is excellent. The computational effort of this optimization
scales as  $D^4 d^2$ in memory and similarly in time (for a single
site optimization).

But what about the theoretical worst case computational complexity
of finding this optimal MPS? It has been observed that DMRG
converges exponentially fast to the ground state with a relaxation
time proportional to the inverse of the gap $\Delta$ of the system
\cite{DMRG1}. For translationally invariant critical systems, this
gap seems to close only polynomially. As we have proven that $D$
only have to scale polynomially too, the computational effort for
finding ground states of 1-D quantum systems is polynomial ($P$).
This statement is true under the following conditions: 1) the
$\alpha$-entropy of blocks in the exact ground state grow at most
logarithmically with the size of the block for some $\alpha<1$; 2)
the gap of the system scales at most polynomially with the system
size; 3) given a gap that obeys condition 2, there exists an
efficient DMRG-like algorithm that converges to the global
minimum. As the variational MPS approach \cite{VPC04} is
essentially an alternating least squares method of solving a
non-convex problem, there is a priori no guarantee that it will
converge to the global optimum, although the occurrence of local
minima seems to be  unlikely \cite{DMRG1}. But still, this has not
been proven and the worst-case complexity could well be NP-hard,
as multiquadratic cost functions have been shown to lead to
NP-hard problems \cite{Nemirovski}. \footnote{In fact, it can be
shown \cite{Eisert} that a variant of DMRG leads to NP-hard
instances in intermediate steps. However, one has to note that (i)
such a worst case instance might be avoided by starting from a
different initial point and (ii) convergence to the optimum at the
end of the day does not necessarily require finding the optimum in
every intermediate step.}

\subsection{Classical simulation of quantum circuits}

In the standard circuit model of quantum computation a set of
unitary one and two-qubit gates is applied to a number of qubits,
which are initially in a pure product state and measured
separately in the end. In the cluster state model a multipartite
state is prepared in the beginning and the computation is
performed by applying subsequent single-qubit measurements. For
both computational models the MPS formalism provides a simple way
of understanding why a large amount of entanglement is crucial for
obtaining an exponential speed-up with respect to classical
computations. The fact that quantum computations (of the mentioned
type) which contain too little entanglement can be simulated
classically is a simple consequence of the following observations
(see also \cite{Jozsa}):
\begin{enumerate}
    \item Let $|\psi\rangle\in \mathbb{C}^{\otimes d^N}$ have a MPS representation
    with maximal bond dimension $D$. By Eq.(\ref{E}) the expectation values of factorizing
    observables are determined by a product of $N$ $D^2\times D^2$
    matrices. Hence, their calculation as well as the sampling of the respective
    measurement outcomes and the storage of
    the state requires only polynomial resources in $N$ and $D$.
    \item A gate applied to two neighboring sites increases the
    respective bond dimension (Schmidt rank) at most by a factor
    $d^2$ and the MPS representation can be updated with $poly(D)$
    resources.
    \item A gate applied to two qubits, which are $l$ sites apart,
    can be replaced by $2l-1$ gates acting on adjacent qubits. In
    this way we can replace a circuit in which each qubit line is
    crossed or involved by at most $L$ two-qubit gates by one
    where each qubit is affected only by at most $4 L$ nearest
    neighbor gates.
\end{enumerate}
A trivial implication of 1. is that measurement based quantum
computation (such as cluster state computation) requires more than
one spatial dimension:
\begin{theorem}[Simulating 1D measurement based computations] \label{Thm:ClusterComputation}
Consider a sequence of states with increasing particle number $N$
for which the MPS representation has maximal bond dimension
$D=O(poly(N))$. Then every measurement based computation on these
states can be simulated classically with $O(poly(N))$ resources.
\end{theorem}
This is in particular true for a one-dimensional cluster state,
for which $D=2$ is independent of $N$. Similarly, if the initial
states are build by a constant number of nearest neighbor
interactions on a two-dimensional $n\times \log n$ lattice, the
measurements can be simulated with $poly(n)$ resources.

Concerning the standard circuit model for quantum computation on
pure states the points 1.-3. lead to:
\begin{theorem}[Simulating quantum circuits]\label{Thm:CircuitComputation}
If at every stage the $N$-partite state in a polynomial time
circuit quantum computation has a MPS representation with
$D=O(poly(N))$, then the computation can be simulated classically
with $poly(N)$ resources. This is in particular true if every
qubit-line in the circuit is crossed or involved by at most $O(log
N)$ two-qubit gates.
\end{theorem}
For more details on the classical simulation of quantum circuits
we refer to \cite{Jozsa} and \cite{YoranShort} for measurement
based computation as well as for the relation between contracting
tensor networks and the tree width of network graphs. The general
problem of contracting tensor networks was shown to be $\sharp P$
complete in \cite{PEPScomplexity}.
 In
\cite{FCcluster} a simple interpretation of measurement based
quantum computation in terms of the valence bond formalism is
provided, which led to new resources in \cite{newresources}.

\section*{Acknowledgments}

D. P-G was partially supported by Spanish projects ``Ram\'on y
Cajal" and MTM2005-00082.

\appendix

\section{An open problem}\label{open.problems}

The main open problem we left open is the {\it reasonability} of
assuming condition C1 in the blocks of the canonical form. Since
condition C2 is known to imply C1 for sufficiently large $L_0$
\cite{FaNaWe92} and we do have C2 in each block by construction,
it only remains to show how {\it big} $L_0$ can be. (We are
neglecting complex eigenvalues of $\E$ of modulus $1$. The reason
is that, by Thm.\ref{Th:periodic}, they can be avoided by grouping
the spins into blocks or simply considering $N$ prime.) In the
generic case this $L_0$ can be taken
$L_0=\lfloor\frac{2\ln{D}}{\ln{d}}\rfloor+1$. However there are
cases in which $L_0\sim O(D^2)$. For instance
$A_0=\sum_{i=1}^D|i+1\>\<i|$, $A_1=|2\>\<D|$. Our conjecture is
that this is exactly the worst case.

\begin{conj}\label{Conj1}
There exists a function $f(D)$ such that for any sequence of
$D\times D$ matrices $A_i$ verifying C1 for some $L_0$, $L_0$ can
be taken $f(D)$.
\end{conj}

\begin{conj}\label{Conj2}
$f(D)\sim O(D^2)$.
\end{conj}

We have been able to prove both in a particular (but generic)
case:

\begin{prop}\label{prop:appendix}
If $A_0$ is invertible, then $L_0$ can be taken $D^2$.
\end{prop}

\proof{ Take an $r$ and call $S_r\subset \{0,\ldots,d-1\}^r$ to a
maximal set of indices for which $\{A_{i_1}\cdots
A_{i_r}\}_{(i_1,\ldots,i_r)\in S}$ is linearly independent. It is
then sufficient to prove that the cardinalities $\#S_{r+1}>\#S_r$,
whenever $\#S_r<D^2$ . If not, using that $A_0$ is invertible, we
can assume that $S_{r+1}=S_r\times\{0\}$. Now $A_{i_1}\cdots
A_{i_{r+2}}=$
$$A_{i_1}\sum_{(j_2,\ldots,j_{r+1})\in S_r}
c^{j_2,\ldots,j_{r+1}}_{i_2,\ldots, i_{r+2}}A_{j_{2}}\cdots
A_{j_{r+1}}A_0=$$
$$\sum_{(j_2,\ldots,j_{r+1})\in S_r} c^{j_2,\ldots,j_{r+1}}_{i_2,\ldots,
i_{r+2}}\sum_{(k_1,\ldots,k_{r})\in S_r}
c^{k_1,\ldots,k_{r}}_{i_1,j_2\ldots, j_{r+1}}A_{k_1}\cdots A_{k_r}
A_0^2,$$ which implies that one can take
$S_{r+2}=S_r\times\{0\}\times\{0\}$ and hence
$\#S_{r+2}=\#S_{r+1}=\#S_r$. We can continue the reasoning and
prove that, in fact, $\#S_{k}=\#S_r$ for every $k\ge r$; which is
the desired contradiction. }

A particular case of this proposition occurs when one of the
matrices in the canonical decomposition is hermitian. To see that
one can group the spins in blocks of two. The new matrices are
$\tilde{A}_{ij}=A_iA_j$, and the canonical condition $\sum_i
A_i^2=\1$ reads $\sum_i \tilde{A}_{ii}=1$. Now, by doing a unitary
operation in the physical system on can assume that
$\tilde{A}_0=\sum_i \tilde{A}_{ii} =\1$ (up to normalization) and
therefore we are in the conditions of Proposition
\ref{prop:appendix}.

\

The conjectures, if true, can be used to prove a couple of
interesting results, one concerning the MPS representation of the
W-state, and the other concerning the approximation by MPS of
ground states of gapped hamiltonians

\subsection{W state}

The $W$-state can be written with matrices $2\times 2$ in the OBC
representation. However, this representation does not respect the
symmetry of the state. What if we ask instead for representations
of this kind?

Let us start with a permutational invariant one, that is, one with
diagonal matrices. By the permutational invariance of $|W_N\>$, it
can be written as a sum $\sum_{n=1}^{N+1} \delta_n |x_n\>^{\otimes
N}$. Calling $\alpha_n$ to a $N$-th root of $\delta_n$ and
$A_i=\sum_n \alpha_n \<i|x_n\> |n\>\<n|$,
$$|W_N\>=\sum_{i_1,\ldots, i_N=1}^2 \trace(A_{i_1}\cdots A_{i_N})|i_1\cdots i_N\>.$$
with the matrices $A_i$ diagonal and of size $(N+1)\times (N+1)$.

\

Of course we would be interested in a representation with smaller
matrices. The surprising consequence of Conjecture \ref{Conj2} is
that this is impossible even if we ask for a TI representation;
that is, one with site-independent matrices (not necessarily
diagonal):

\begin{corollary}
If we assume Conjecture \ref{Conj2} and $|W_N\>=\sum_{i_1,\ldots,
i_N=1}^2 \trace(A_{i_1}\cdots A_{i_N})|i_1\cdots i_N\>$ for
$D\times D$ matrices $A_i$, then $D\succeq O\left(N^{1/3}\right)$.
\end{corollary}

\proof{ By Conjecture \ref{Conj2}, we have condition C1 by blocks
with $L_0=O(D^2)$. Let us assume that $\frac{N}{2}\ge
3(b-1)(L_0+1)$, where (as usual) $b$ is the number of different
blocks in the canonical form. We can break the chain in two parts,
each one having at least $3(b-1)(L_0+1)$ spins and decompose the
state as $|W_N\>=\sum_{\alpha,\beta=1}^D |\Phi_{\alpha,\beta}\>
|\Psi_{\alpha,\beta}\>,$ where
$$|\Phi_{\alpha,\beta}\>=\sum_{i_1,\ldots,i_R}
\<\alpha|A_{i_1}\cdots A_{i_R}|\beta\>|i_1\cdots i_R\rangle$$
$$|\Psi_{\alpha,\beta}\>=\sum_{i_1,\ldots,i_R}
\<\beta|A_{i_R}\cdots A_{i_N}|\alpha\>|i_{R+1}\cdots i_N\rangle.$$

If we call $S_j$ to the set of positions of the $j$-th block in
the canonical form of $|W_N\>$ and $S=\cup S_j$, we claim that the
sets $\{|\Phi_{\alpha,\beta}\>\}_{(\alpha,\beta)\in S}$ and
$\{|\Phi_{\alpha,\beta}\>\}_{(\alpha,\beta)\in S}$ are both
linearly independent. This implies that the rank of the reduced
density operator after tracing out the particles $1,\ldots, R$ is
$\ge \sum_{i=1}^bD_i^2$, being $D_i\times D_i$ the size of the
$i$-th block.

We know that in the case of $|W_N\>$, this rank is $2$, so we get
exactly two blocks of size $1\times 1$. It is trivial now to see
that this is impossible. Therefore, $\frac{N}{2}\le 3(b-1)(L_0+1)$
and hence the result.

\

So it only remains to prove the claim. For $(\alpha,\beta)\in S_j$
let us take complex numbers $c_{\alpha,\beta}$ such that
$\sum_{\alpha,\beta} c_{\alpha,\beta} |\Phi_{\alpha,\beta}\>=0$ ;
which is exactly
$$\sum_{j=1}^b\trace(\left[\sum_{(\alpha,\beta)\in S_j}c_{\alpha,\beta}|\beta\>\<\alpha|\right]A^j_{i_1}\cdots
A^j_{i_R})|i_1\cdots i_R\>=0.$$ By Lemma \ref{lem:direct-sum} the
sum in $j$ is direct, so each summand is $0$. Finally, by
condition C1 by blocks  $\sum_{(\alpha,\beta)\in
S_j}c_{\alpha,\beta}|\beta\>\<\alpha|=0$ for every $j$ and hence
$c_{\alpha,\beta}=0$ for every $(\alpha,\beta)\in S_j$. }

\subsection{Approximation of ground states by MPS}

As we saw in Sec.\ref{sec:classical-sim}, one of the big open
questions in condensed matter theory is to mathematically explain
the high accuracy of DMRG. Since DMRG can be seen as a variational
method in the class of MPS, it is crucial to prove that any ground
state of a gapped local Hamiltonian can be efficiently
approximated by a MPS of low bond dimension $D$. Despite the
important recent advances in this direction (see
Sec.\ref{sec:classical-sim}) the problem is still unsolved. An
important step has been done by Hastigns, who has recently reduced
this problem to the case in which the Hamiltonian is frustration
free \cite{Hast1}. This highlights the importance of the following
dichotomy

\begin{theorem}[Dichotomy for the size of the MPS]

If Conjecture \ref{Conj2} holds and $H_N=\sum_{i=1}^N \tau^i(h)$
is a local TI Hamiltonian which is frustration free for every $N$,
then the bond dimension $D$ of any of its {\it exact} ground
states, viewed as a MPS, is:

\begin{enumerate}
\item[(i)] either independent of $N$ \item[(ii)] or
$>O(N^{\frac{1}{3}})$
\end{enumerate}

\end{theorem}

\proof{ Let us take the canonical decomposition of a ground state
$|\psi\>$ of $H$ acting on $N$ particles
$$|\psi\>=\sum_{i_1,\ldots, i_N=1}^d \trace(A_{i_1}\cdots
A_{i_N})|i_1\cdots i_N\>.$$ If (ii) does not hold, by Conjecture
\ref{Conj2}, $N>3(b-1)(L_0+1)+L$ ($L$ the interaction length of
$H$). Now, by Lemma \ref{lem:direct-sum}, the products of the last
$N-L$ matrices generate the whole space of block diagonal
matrices. This immediately implies that the same matrices $A_i$
give us a ground state of $H_{N'}$ for any $N'> N$. That is, $D$
is independent of $N$. }


\begin{thebibliography}{99}

\bibitem{FaNaWe92} M. Fannes, B.
Nachtergaele and R. F. Werner, Commun. Math. Phys. \textbf{144},
443-490 (1992)

\bibitem{MPSorigin} A. Kl\"umper, A. Schadschneider, J. Zittartz, J. Phys. A {\bf 24},
L955 (1991); Z. Phys. B {\bf 87}, 281 (1992).

\bibitem{AKLT}
I. Affleck, T. Kennedy, E. H. Lieb and H. Tasaki, Commun. Math.
Phys. \textbf{115}, 477 (1988)

\bibitem{FaNaWe92-2}
M. Fannes, B. Nachtergaele and R. F. Werner, Lett. Math. Phys.
\textbf{25}, 249-258 (1992)

\bibitem{Hast1} M. Hastings, Phys. Rev. B {\bf 73}, 085115 (2006).

\bibitem{NachtergaeleSym} B. Nachtergaele, Commun. Math. Phys., {\bf 175}, 565
(1996).

\bibitem{Vi03}
G. Vidal, Phys. Rev. Lett. \textbf{91}, 147902 (2003)

\bibitem{EvansH-Krohn78}
D. E. Evans and E. Hoegh-Krohn, J. London Math. Soc. \textbf{17},
345-355 (1978)

\bibitem{renorm}
F. Verstraete, J.I. Cirac, J.I. Latorre, E. Rico and M.M. Wolf,
Phys.Rev.Lett. \textbf{94}, 140601 (2005)

\bibitem{diagonal-krauss} C. King, quant-ph/0412046; I. Devetak, P.W. Shor,
quant-ph/0311131; M.M. Wolf, D. P\'{e}rez-Garc{\'{\i}}a, Phys.
Rev. A {\bf 75}, 012303 (2007).


\bibitem{HornJohnson91}
R.A. Horn and C.R. Johnson, \emph{Topics in Matrix Analysis},
Cambridge University Press, 1991.


\bibitem{Knabe}
S. Knabe, J. Stat. Phys.  \textbf{52}, 627 (1988).

%---------------------------
\bibitem{SSVCW} C. Schoen, E. Solano, F. Verstraete, J. I. Cirac, M. M.
Wolf, Phys. Rev. Lett. {\bf 95}, 110503 (2005).
\bibitem{SSVCW2} C. Schoen, K. Hammerer, M.M. Wolf, J.I. Cirac, E. Solano
, Phys. Rev. A {\bf 75}, 032311 (2007).
\bibitem{Kike} Y. Delgado, L. Lamata, J. Leon, D. Salgado, E.
Solano, Phys. Rev. Lett. {\bf 98}, 150502 (2007).
\bibitem{entropyMPS}  N. Schuch, M.M. Wolf, F. Verstraete, J.I. Cirac,
arXiv:0705.0292 (2007).
\bibitem{Wilson} K.G. Wilson, Rev. Mod. Phys. {\bf 47}, 773 (1975).
\bibitem{White} S. White, Phys. Rev. Lett. \textbf{69}, 2863 (1992)
\bibitem{Ostlund} S. Ostlund and S. Rommer, Phys. Rev. Lett. \textbf{75}, 3537 (1995); S. Rommer and S. Ostlund, Phys. Rev. B \textbf{55}, 2164 (1997)
\bibitem{vonDelft} F. Verstraete, A. Weichselbaum, U. Schollw\"{o}ck, J. I. Cirac, Jan von Delft, cond-mat/0504305.
\bibitem{VPC04} F. Verstraete, D. Porras and I. Cirac, Phys. Rev. Lett. \textbf{93}, 227205 (2004).
\bibitem{Bombelli-Srednicki-Bousso} L. Bombelli, R.K. Koul, J. Lee, R.D. Sorkin, Phys. Rev. D {\bf 34},
373 (1986); M. Srednicki, Phys. Rev. Lett. {\bf 71}, 66 (1993); R.
Bousso, Rev. Mod. Phys. {\bf 74}, 825 (2002).
\bibitem{arealaw} G. Vidal, J. I. Latorre, E. Rico, A. Kitaev, Phys. Rev. Lett. {\bf 90},
227902 (2003); B.Q. Jin, V. Korepin, J.Stat.Phys. {\bf 116}, 79
(2004); P. Calabrese, J. Cardy, J.Stat.Mech. P06002 (2004); M.B.
Plenio  J. Eisert, J. Dreissig, M. Cramer, Phys. Rev. Lett. {\bf
94}, 060503 (2005); M.M. Wolf, Phys. Rev. Lett. {\bf 96}, 010404
(2006); D. Gioev and I. Klich, Phys. Rev. Lett. {\bf 96}, 100503
(2006); F. Verstraete, M.M. Wolf, D. Perez-Garcia, J.I. Cirac,
Phys. Rev. Lett. {\bf 96}, 220601 (2006); M.M. Wolf, F.
Verstraete, M.B. Hastings, J.I. Cirac, arXiv:0704.3906 (2007).
\bibitem{Ortiz} M.M. Wolf, G. Ortiz, F. Verstraete, J.I. Cirac, Phys. Rev. Lett. {\bf 97}, 110403 (2006).
\bibitem{Cardy} P. Calabrese and J. Cardy, J. Stat. Mech. (2004)
P06002.
\bibitem{Jin} B.-Q. Jin and  V.E. Korepin,  J. Stat. Phys. {\bf 116}, 79
(2004).
\bibitem{hastings1} M. B. Hastings, Phys. Rev. Lett. {\bf 93}, 140402 (2004);
M.B. Hastings, T. Koma, Commun. Math. Phys. {\bf 265}, 781 (2006);
B. Nachtergaele, R. Sims, Commun. Math. Phys., {\bf 265}, 119
(2006).
\bibitem{datahiding} P. Hayden, D. Leung, P. W. Shor, A. Winter, Commun. Math.
Phys. {\bf 250}, 371 (2004).
\bibitem{VeCi05} F. Verstraete and I. Cirac, Phys. Rev. B
\textbf{73}, 094423 (2006)
\bibitem{DMRG1} U. Schollw\"{o}ck, Rev. Mod. Phys. {\bf 77}, 259 (2005).
\bibitem{Nemirovski}  A. Ben-Tal and A.Nemirovski, Mathematics of Operational Research {\bf 23}, 769 (1998).
\bibitem{Eisert} J. Eisert, Phys. Rev. Lett. {\bf 97}, 260501
(2006).
\bibitem{Jozsa} R. Jozsa, quant-ph/0603163 (2006); G. Vidal, Phys. Rev. Lett. {\bf 91},
147902 (2003); D. Aharonov, Z. Landau, J. Makowsky,
quant-ph/0611156.
\bibitem{YoranShort} N. Yoran, A. Short, quant-ph/0601178 (2006); M. Van den Nest, W. D\"{u}r, G. Vidal, H. J. Briegel,
Phys. Rev. A {\bf 75}, 012337 (2007); I. Markov, Y. Shi,
quant-ph/0511069 (2005).
\bibitem{PEPScomplexity} N. Schuch, M.M. Wolf, F. Verstraete, J.I. Cirac, Phys. Rev. Lett. {\bf 98}, 140506
(2007).
\bibitem{FCcluster} F. Verstraete, J.I. Cirac, Phys. Rev. A {\bf
70}, 060302(R) (2004).
\bibitem{newresources} D. Gross, J.Eisert, arXiv:quant-ph/0609149
(2006).

\end{thebibliography}
\end{document}